\documentclass[12pt]{article}

\usepackage{amssymb}
\usepackage{amsmath}
\usepackage{latexsym}
\usepackage{yfonts}

\catcode`@=11
\renewcommand{\section}{\@startsection{section}{1}{0pt}{\medskipamount}
{\medskipamount}{\large\bf}} \numberwithin{equation}{section}
\catcode`@=12

\def\beq{\begin{eqnarray}}    
\def\eeq{\end{eqnarray}}      

\def\ln{\,\mbox{ln}\,}                  
\def\Tr{\,\mbox{Tr}\,}                  

\def\pa{\partial}                       

\def\={\ =\ }

\def\al{\alpha}
\def\be{\beta}

\def\ga{\gamma}
\def\de{\delta}

\def\vp{\varepsilon}

\def\om{\omega}

\def\Ga{\Gamma}

\def\La{\Lambda}

\begin{document}

\begin{center}

{\Large\bf BRST, Ward identities, gauge dependence, and a functional
renormalization group}

\vspace{18mm}

{\large P.M. Lavrov$^{(a, b)}\footnote{E-mail:
lavrov@tspu.edu.ru}$\; }

\vspace{8mm}

\noindent  ${{}^{(a)}} ${\em
Tomsk State Pedagogical University,\\
Kievskaya Street.\ 60, 634061 Tomsk, Russia}

\noindent  ${{}^{(b)}} ${\em
National Research Tomsk State  University,\\
Lenin Avenue\ 36, 634050 Tomsk, Russia}

\vspace{20mm}

\begin{abstract}
\noindent
Basic properties of gauge theories in the framework of
Faddeev-Popov (FP) method, Batalin-Vilkovisky (BV) formalism, functional
renormalization group (FRG) approach are considered. The FP and BV
quantizations are characterized by the Becchi-Rouet-Stora-Tyutin
(BRST) symmetry while the BRST
symmetry is broken in the FRG approach. It is shown that the
FP method, the BV formalism and the FRG approach can be provided
with the Slavnov-Taylor identity, the Ward identity and the modified
Slavnov-Taylor identity, respectively. It is proven that using the
background field method  the background gauge invariance of
effective action within the FP and FRG  quantization procedures can
be achieved in nonlinear gauges. The gauge-dependence problem
within the FP, BV and FRG quantizations is studied.  Arguments allowing us
to state the existence of principal problems of the FRG in the case
of gauge theories are given.

\end{abstract}

\end{center}

\vfill

\noindent {\sl Keywords:} BRST symmetry, Ward identities, gauge
dependence, functional renormalization group
\\

\noindent PACS numbers: 11.10.Ef, 11.15.Bt
\newpage

\section{Introduction}
\noindent
Over the past three decades, there has been an increased interest in the
nonperturbative approach in quantum field theory known
as the functional renormalization group (FRG), which has been proposed
in papers \cite{Wet1,Wet2} and  can be considered as a
version of Wilson renormalization group  \cite{Wilson,Polch}.
The FRG approach  has gotten further
developments \cite{Wett-Reu-1,FRG7,RW1,Ell,Bob-Att-Mar-1,Bob-Att-Mar-2,Att-TimM}
and numerous applications
\cite{Bob-Att-Mar-3,Ell-H-W,ReWe-1997,Lit-Paw-1,Fre-Lit-Paw,
BG,Iga-Ito-So-1,Iga-Ito-So-2,LPaw,BR,NPS,CFPawR}. There are many reviews  devoted
to detailed discussions of different aspects of the FRG approach
and among them one can find Refs.
\cite{FRG2,FRG1,FRG3,FRG4,IIS-2009,FRG5,FRG6,Giess} with qualitative references.

As a quantization procedure the FRG  belongs to covariant quantization schemes.
In the case of gauge theories, any covariant quantization  faces
 two principal problems: the
unitarity of S-matrix first formulated by Feynman \cite{Feynman}
and the gauge dependence of results obtained.
The study of the unitarity problem requires
consideration of canonical formulation of a given theory
on the quantum level and use of the Kugo-Ojima method
in construction and analysis of physical state  space with the help of
nilpotent Becchi-Rouet-Stora-Tyutin (BRST) operator \cite{KO}
to discovery the criteria providing the unitarity.
In the present paper, we will not touch the unitary problem in all covariant
quantization
approaches to gauge theories, restricting ourselves the gauge dependence problem.

The gauge dependence is a problem in the quantum description of gauge
theories beginning with famous papers by Jackiw \cite{J}  and
Nielsen \cite{Niel}. Study of the gauge dependence problem can be
directly performed in covariant quantization schemes, namely, in the
Faddeev-Popov (FP) method \cite{FP}, the Batalin-Vilkovisky (BV)
formalism \cite{BV,BV1} and the FRG approach \cite{Wet1,Wet2}).
Analysis of the gauge dependence problem for Yang-Mills theories
in the framework of the FP-method and for general gauge theories
 within the BV-formalism
has been given in papers \cite {LT3,LT1} and \cite{VLT}, respectively.
 Aspects of gauge invariance and related
topics were always under close attention in the FRG
\cite{Ell,Att-TimM,Bob-Att-Mar-3,Fre-Lit-Paw,
BG,Iga-Ito-So-2,C,Wet-2018,deAlwis,Morris1,Morris2,AGZ}.
Nevertheless, it seems useful and important task  to consider the
gauge-dependence problem within the FRG approach for different types
of gauge theories from general points of view.

We are going to compare with each other basic properties providing the FP method,
BV formalism and the FRG approach and  find new features concerning
the gauge dependence problem in the FRG. Among the basic properties,
it needs first of all to mention the BRST symmetry \cite{brs1,t}, which
is considered  a fundamental principle of modern quantum
field theory allowing a suitable quantum description
of a given dynamical system \cite{Weinberg,Green}.
For the first time, the BRST symmetry was discovered
as a global supersymmetry of quantum action (the Faddeev-Popov action)
appearing in the process of quantization of Yang-Mills theories.
In its turn, the BRST symmetry in the BV formalism  is not  the global
supersymmetry of some action, but it is encoded into the quantum master equation.
The role of BRST symmetry in the FP method and in the BV formalism
is extremely important because it guarantees the gauge independence  of
the S-matrix   elements. The BRST symmetry is broken in the FRG
approach, which leads to the ill-defined S-matrix \cite{LSh}.

The Ward identities in quantum theory of gauge fields
are the next basic property. Their existence is a direct consequence of gauge
invariance of the initial classical action underlying a given system with gauge
freedom.
The BRST transformations help to present the Ward identities in
a unique  form that sometimes causes incorrect conclusions
concerning relations between the BRST symmetry and the Ward identities;
namely, the Ward identities by themselves do not mean  the existence of
the BRST symmetry for a given gauge system. It is exactly the case of
the FRG approach when it cannot be provided by the BRST symmetry
in presence of the modified Slavnov-Taylor (mST) identities.

In our investigation, we pay special attention to the gauge-dependence
problem within the FP method, the BV formalism and the FRG approach with
or without using the background field method (BFM) \cite{DeW,AFS,Abbott}
because of  its importance
for the physical interpretation of used schemes of quantization.
Our interest in the background field method is caused
by an important property of gauge invariance of the background
effective action under gauge transformations of background fields helping
to simplify quantum calculations in the Yang-Mills and gravity
theories within the FP method. Unfortunately, this method does not help
to improve the situation with the gauge-dependence problem in the FRG
because the effective
average action being a gauge-invariant functional remains a gauge dependent object.

The paper is organized as follows. In Sec. II, a brief description of
theories invariant under the gauge transformations from the point of view
the structure of corresponding gauge algebras is given. In Sec. III, the BRST
symmetry in the context of FP method, BV formalism and  FRG approach
is discussed. In Sec. IV, the Slavnov-Taylor (ST) identity in the FP method,
the Ward identity
in BV ormalism and the mST identity in FRG approach are studied. In Sec. V,
the gauge-dependence problem is studied within quantization schemes mentioned
above. In Sec. VI, the  all basic properties of FP method and FRG approach
are investigated for the Yang-Mills type of gauge theories within the BFM.
Finally, in Sec. VII, the results obtained in the paper are discussed.

We use the DeWitt's condensed notations \cite{DeWitt}.
We employ the notation $\varepsilon(A)$ for the Grassmann parity of
any quantity $A$.  The right and left functional derivatives with respect to
fields and antifields are marked by special symbols $"\leftarrow"$  and
$"\rightarrow"$, respectively.
Arguments of any functional are enclosed in square brackets
$[\;]$,
and arguments of any function are enclosed in parentheses, $(\;)$.
The symbol $F_{,A}[\phi,...]$ means the
right derivative of $F[\phi,...]$ with respect to field $\phi^A$.

\section{Gauge theories}
\noindent
Let us start from some initial classical action $S_0[A]$ of
the fields $A^i$,
with Grassmann parities $\varepsilon(A^i)\equiv\varepsilon_i$,
being invariant under the gauge transformations
($X_{,}\equiv\delta X/\delta A^i$)
\begin{eqnarray}
\label{a1}
\delta A^i=R^i_{\alpha}(A)\xi^{\alpha},\quad
S_{0,i}[A]R^i_{\alpha}(A) =0,
\end{eqnarray}
where $\xi^{\alpha}$ are arbitrary functions with Grassmann parities
$\varepsilon(\xi^{\alpha})\equiv\varepsilon_{\alpha}$,
$\alpha=1,2,...,m$, and $R^i_{\alpha}(A)$,
$\varepsilon(R^i_{\alpha}(A))=\varepsilon_i + \varepsilon_{\alpha}$
are generators of gauge transformations. It is assumed the set of
fields $A^i$ is linear independent (in particular, it is not the
case of higher-spin fields \cite{Fronsdal}). The general form of
algebra of generators $R^i_{\alpha}(A)$ reads
\begin{eqnarray}
\label{GAGGT}
R^i_{\alpha , j}(A)R^j_{\beta}(A)-(-1)^{\varepsilon_{\alpha}\varepsilon_
{\beta}}R^i_{\beta ,j}(A)R^j_{\alpha}(A)
=-R^i_{\gamma}(A)F^{\gamma}_{\alpha\beta}(A)-
S_{0,j}[A]M^{ij}_{\alpha\beta}(A),
\end{eqnarray}
where $F^{\gamma}_{\alpha\beta}(A)
=-(-1)^{\varepsilon_{\alpha}\varepsilon_{\beta}}
F^{\gamma}_{\beta\alpha}(A)$ are  structure functions depending, in general,
on the fields $A^i$ and $M^{ij}_{\alpha\beta}(A)$ satisfies the conditions
$M^{ij}_{\alpha\beta}(A) = -(-1)^{\varepsilon_i\varepsilon_j}
M^{ji}_{\alpha\beta}(A) =
-(-1)^{\varepsilon_{\alpha}\varepsilon_{\beta}}M^{ij}_{\beta\alpha}(A)$.

If the structure functions do not depend on fields $A^i$,
$M^{ij}_{\alpha\beta}(A)=0$ and, in addition, the generators $R^i_{\alpha}(A)$
form a set of linear independent operators with respect to the index $\alpha$,
then
we have the case of the {\it Yang-Mills type of gauge theories} being very important
for practical
applications because all modern models of fundamental forces are described
in terms of such a kind of theories.

For an example, let
us consider the case of the pure Yang-Mills theory, defined by the
action
\beq
\label{ActionYM}
S_{YM} [A] = - \frac{1}{4} F^a_{\mu\nu}(A) F^a_{\mu\nu} (A),
\eeq
where
$\,F^a_{\mu\nu}(A) = \pa_\mu A^a_\nu - \pa_\nu A^a_\mu
+ f^{abc} A^b_\mu A^c_\nu\,$
is the field strength for the non-Abelian  vector field $A_\mu$,
taking values in the adjoint representation of  a compact semisimple
Lie group with structure coefficients $f^{abc}$. We have the
following identifications with previous notations
\beq
\label{a4}
A^i \mapsto A^a_\mu ,
\qquad
F^{\al}_{\be\ga} \mapsto f^{abc} ,
\qquad R^i_\al(A) \mapsto D^{a b}_\mu(A)
= \de^{ab} \partial_\mu + f^{acb} A^c_\mu.
\eeq
Here, $D^{a b}_\mu(A)$ is the covariant derivative.

For a second example, consider the case of quantum gravity theories,
defined by an
action $S_{0} (g)$ of a Riemann metric $g = \lbrace g_{\mu\nu}(x)\rbrace$
with $\vp(g)=0$,\footnote{The standard example
is Einstein gravity with a cosmological constant term,
\beq
\nonumber
S_{0}[g]
&=& - \, \frac{1}{\kappa^2}
\int dx \sqrt{-{\rm det}g}\, \big(R(g) + 2\La\big).
\eeq
}
and which is invariant under general coordinate transformations.
The generator of such transformation is linear in $g_{\mu\nu}$ and reads
\beq
\label{a5}
R_{\mu\nu\sigma}(x,y;g) = - \delta(x-y) \partial_\sigma g_{\mu\nu}(x) -
g_{\mu\sigma}(x) \partial_\nu \delta(x-y) - g_{\sigma\nu}(x)
\partial_\mu \delta(x-y).
\eeq
Therefore, for an arbitrary gauge function $\xi^\alpha$ with
$\vp(\xi^\alpha) =0$,
one has $\delta g_{\mu\nu} = R_{\mu\nu\sigma}(g) \xi^\sigma$, or,
writing all the arguments explicitly,
\beq
\label{a6}
\delta g_{\mu\nu}(x) =
\int dy \, R_{\mu\nu\sigma}(x,y;g) \xi^\sigma(y).
\eeq
In this case, the structure functions are given by
\beq
\label{a7}
F^{\al}_{\be\ga}(x,y,z) = \delta(x-y)\delta^\alpha_\gamma
\partial^{(x)}_\be \delta(x-z)
 - \delta(x-z) \delta^\al_\be \partial^{(x)}_\ga \delta(x-y),
\eeq
which satisfy the antisymmetry properties,
$F^{\al}_{\be\ga}(x,y,z) = - F^{\al}_{\ga\be}(x,z,y)$, as usual.

In terms of the notation used,  one has the correspondence
\beq
\label{a8}
A^i \mapsto g_{\mu\nu}(x) ,
\qquad R^i_\al(A) \mapsto R_{\mu\nu\sigma}(x,y;g) ,
\qquad
F^{\al}_{\be\ga} \mapsto F^{\al}_{\be\ga}(x,y,z).
\eeq

In general, the structure functions may depend on fields $A^i$,
$M^{ij}_{\alpha\beta}(A)$ may not be equal to zero ({\it open algebras}),
and $R^i_{\alpha}(A)$ may not be linear independent in the index $\alpha$
({\it reducible algebras}). In all these cases, we meet the so-called
{\it general gauge theories} \cite{BV,BV1}. For our goals, a detailed description
of structure of gauge algebras is not essential,
and we omit their further discussions.

All results obtained below within the FP method and the FRG are valid for any
Yang-Mills type of gauge theories in any admissible gauge. The same remark is valid
for general gauge theories in the BV formalism.

\section{BRST symmetry}
\noindent
At present, the BRST symmetry is considered as a fundamental principle
in the construction of the consistent quantization procedure for field and string
theories  \cite{Weinberg,Green}. In the next three subsections, we are
going to discuss a status
of the BRST symmetry for the Yang-Mills type of gauge theories within
the FP method and the FRG approach and for the general gauge
theories  within the BV formalism.

\subsection{BRST  in FP-method}
\noindent
Let $S_0[A]$ be an action of fields $A^i$ which include Yang-Mills fields and,
in general,  multiplets of spinor and scalar fields.
Vacuum functional for Yang-Mills type of gauge theories is
constructed by the Faddeev-Popov rules \cite{FP} in the form
of functional integral
\beq
\label{b1}
Z=\int D\phi \;\exp\Big\{\frac{i}{\hbar}S_{FP}[\phi]\Big\}
\eeq
over  fields $\phi$.
In (\ref{b1}),
$S_{FP}[\phi]$ is the Faddeev-Popov action,
\beq
\label{b2}
S_{FP}[\phi]=S_{0}[A]+
{\bar C}^{\alpha}\left(\chi_{\alpha}(A,B)
\overleftarrow{\pa}_{\!\!A^i}\right)R^i_{\beta}(A)\;\!C^{\beta}+
B^{\alpha}\chi_{\alpha}(A,B),
\eeq
where $\chi_{\alpha}(A,B)$ are functions lifting the degeneracy
of the Yang-Mills action,
$\phi=\{\phi^A\}$  is the set of all fields
\beq
\label{b3}
\phi^A=(A^i,B^{\alpha}, C^{\alpha}, {\bar C}^{\alpha}),\quad
\varepsilon(\phi^A)=\varepsilon_A,
\eeq
with the Faddeev-Popov ghost and anti-ghost fields
$ C^{\alpha}, {\bar C}^{\alpha}$ ($\varepsilon(C^{\alpha})=
\varepsilon({\bar C}^{\alpha})=1,\;
{\rm gh}(C^{\alpha})=-{\rm gh}({\bar C}^{\alpha})=1$), respectively, and
the Nakanishi-Lautrup auxiliary fields $B^{\alpha}$
($\varepsilon(B^{\alpha})=0,\; {\rm gh}(B^{\alpha})=0$).
A standard choice of linear and nondegenerate gauges $\chi_{\alpha}(A,B)$
reads
\beq
\label{b4}
\chi_{\alpha}(A,{\cal B})=F_{\alpha i}A^i+\frac{\xi}{2}B_{\alpha},
\eeq
where $F_{\alpha i}$, being some differential operations, do not
depend on fields $A^i$  and $\xi$ is a constant gauge parameter.
In what follows we do not restrict ourselves by the case (\ref{b4}) and
consider the gauge-fixing functions in general settings.

The action
 (\ref{b2}) is invariant under global supersymmetry (BRST symmetry)
 \cite{brs1,t}\footnote{For more compact presentation, we
use the notation $\delta_B$ for $\delta_{BRST}$.}
\beq
\label{b5}
\delta_{B} A^i=R^i_{\alpha}(A)C^{\alpha}\mu,\quad
\delta_{B} C^{\alpha}=-\frac{1}{2}(-1)^{\varepsilon_{\beta}}
F^{\alpha}_{\beta\gamma}
C^{\gamma }C^{\beta}\mu,\quad
\delta_{B}\overline{C}^{\alpha}=B^{\alpha}(-1)^{\varepsilon_{\alpha}}
\mu,\quad
\delta_{B} B^{\alpha} =0,
\eeq
where  $\mu$ is a constant anticommuting parameter or, in short,
\beq
\label{b6}
\delta_{B}\phi^A=R^A(\phi)\mu,\quad
\varepsilon(R^A(\phi))=\varepsilon_A+1,
\eeq
where
\beq
\label{b7}
R^A(\phi)=\big(R^i_{\alpha}(A)C^{\alpha},\; 0\;,
-\frac{1}{2}(-1)^{\varepsilon_{\beta}}
F^{\alpha}_{\beta\gamma}
C^{\gamma }C^{\beta}, B^{\alpha}(-1)^{\varepsilon_{\alpha}}\big).
\eeq
Introducing the gauge fixing functional $\Psi=\Psi[\phi]$,
\beq
\label{b8}
\Psi={\bar C}^{\alpha}\chi_{\alpha}(A,B),
\eeq
the action (\ref{a7}) is rewritten in the form
\beq
\label{b9}
S_{FP}[\phi]={S}_{0}[A]+\Psi[\phi]
{\hat R}(\phi)={S}_{0}[A]+\Psi[\phi]_{,A}R^A(\phi),
\qquad S_{0}[A]{\hat R}(\phi)=0,
\eeq
where
\beq
\label{b10}
{\hat R}(\phi)=
\overleftarrow{\pa}_{\!\!\phi^A}R^A(\phi)
\eeq
is the generator of BRST transformations.
Because of the nilpotency property of ${\hat R}$, ${\hat R}^2=0$,
the BRST symmetry of $S_{FP}$
follows from the presentation (\ref{b9}) immediately,
\beq
\label{b11}
S_{FP}[\phi]{\hat R}(\phi)=0.
\eeq

The BRST symmetry of $S_{FP}$ leads to a very important property
of the vacuum functional (\ref{b1}), namely, its gauge independence. Indeed, let
$Z_{\psi}$ be vacuum functional corresponding to choice of gauge-fixing
functional
$\Psi$. Consider the vacuum functional  for another choice of gauge condition
$\Psi+\delta\Psi$, $Z_{\psi+\delta\Psi}$. Then, we have
\beq
\label{b12}
Z_{\Psi+\delta\Psi}=\int D\phi\;\exp\Big\{\frac{i}{\hbar}\big(S_{FP}[\phi]+
\delta\Psi[\phi]{\hat R}(\phi)\big)\Big\}.
\eeq
Making use of change of integration variables in the functional integral (\ref{b12})
in the form of
the BRST transformations (\ref{b6}) but with parameter $\mu$ being
an functional $\mu=\mu[\phi]$ with
\beq
\label{b13}
\mu[\phi]=\frac{i}{\hbar}\delta\Psi[\phi]
\eeq
and taking into account that the Jacobian of the transformations is equal to
\beq
\label{b14}
J=\exp\{-\mu[\phi]{\hat R}(\phi)\},
\eeq
we obtain
\beq
\label{b15}
Z_{\Psi+\delta\Psi}=Z_{\Psi}.
\eeq
In deriving (\ref{b14}), the  relations
\beq
\label{b16}
(-1)^{\varepsilon_i} \overrightarrow{\pa}_{\!\!\!A^i}\;\!R^i_\alpha ( A)
+ (-1)^{\epsilon_\beta + 1} F^\beta_{\beta\alpha}  = 0,
\eeq
were used. In Yang-Mills theories, for instance, the
relations~\eqref{b16} are satisfied due to antisymmetry properties of
the structure constants. The BRST transformations (\ref{b5}) obey the property of
nilpotency, $\delta^2_B\phi^A=0$. In terms of $R^A(\phi)$,  this property means
equalities
\beq
 \label{b17}
R^A_{\;,B}(\phi)R^B(\phi)=0.
\eeq
In turn, the relations  (\ref{b16}) are equivalent to
\beq
\label{b17a}
R^A_{,A}(\phi) =0.
\eeq
We assume the validity of (\ref{b17}) and (\ref{b17a}) in the case
of any Yang-Mills type of gauge theories.

From (\ref{b15}), we conclude the gauge independence of vacuum functional. It was
the reason for us to drop  subscript $\Psi$ in the vacuum functional (\ref{b1}).
The gauge independence of $Z$ is closely related with the BRST symmetry
of $S_{FP}[\phi]$ and leads to the gauge independence of S-matrix elements
due to the equivalence theorem \cite{KT}.

\subsection{BRST in BV-formalism}
\noindent
Let $S_0[A]$ be an initial classical action belonging to the set of general gauge
theories described in Sec. 2. Quantization of this gauge theory
can be performed in the BV formalism \cite{BV,BV1}.
The vacuum functional can be presented in the form of functional integral
\begin{eqnarray}
\label{b18}
Z=\int D\phi \; D\phi^*\; d\lambda\; \exp\Big\{\frac{i}{\hbar}\big(S[\phi,\phi^*]
+(\phi^*_A -\Psi[\phi]\overleftarrow{\pa}_{\!\!\!\phi^A})\lambda^A\big)\Big\}
\end{eqnarray}
where $S=S[\phi,\phi^*]$ is an action satisfying the quantum master equation
\begin{eqnarray}
\label{b19}
\frac {1}{2} (S,S)=i\hbar{\Delta}S
\end{eqnarray}
and the boundary condition
\begin{eqnarray}
\label{b20}
S\big|_{\phi^* = \hbar = 0}= S_0[A].
\end{eqnarray}
The total configuration space $\phi=\{\phi^A\}, \varepsilon(\phi^A)=\varepsilon_A$
is introduced. For irreducible theories the set of fields  $\phi^A$ coincides
with (\ref{b3}). For reducible theories, the set of fields $\phi^A$ has more
complicated structure \cite{BV1} and contains main chains of the ghost,
 antighost and auxiliary Nakanishi--Lautrup fields as well
as pyramids of the ghosts for ghosts  and auxiliary fields.
For our goals here, the explicit structure of $\phi^A$
is not important, only its existence sufficient. To each field $\phi^A$
of the total configuration space, one introduces the corresponding antifield $\phi^*_A$.
The statistics of $\phi^*_A$ is opposite to the statistics of the
corresponding fields $\phi^A$, $\varepsilon (\phi^*_A) = \varepsilon_A + 1$.
 In the left-hand side
of (\ref{b19}) on the space of the fields $\phi^A$ and antifields
$\phi^*_A$, the notation of antibracket
\beq
\label{b21}
(F,G)=F\big(\overleftarrow{\pa}_{\!\!\phi^A}\;
\overrightarrow{\pa}_{\!\!\phi^*_A}-
\overleftarrow{\pa}_{\!\!\phi^*_A}
\overrightarrow{\pa}_{\!\!\phi^A}\big)\;\!G
\eeq
is used. In the right-hand side of (\ref{b19}), $\Delta$ means the second-order
functional differential operator
\beq
\label{b22}
\Delta=
(-1)^{\vp_A}\overrightarrow{\pa}_{\!\!\phi^A}\;\!
\overrightarrow{\pa}_{\!\!\phi^*_A},\quad \vp(\Delta)=1,
\eeq
which obeys the nilpotency property
\beq
\label{b23}
\Delta^2=0.
\eeq
Additionally, in (\ref{b18}), the auxiliary fields
$\lambda^A,\;\;\varepsilon(\lambda^A) = \varepsilon_A + 1$ are introduced.
Finally, in (\ref{b18}), $\Psi=\Psi[\phi]$ is suitable
odd gauge-fixing functional.

Note, first of all, that the integrand in (\ref{b18}) is
invariant under the following global supertransformations:
\begin{eqnarray}
\label{b24}
\delta_B\phi^A = \lambda^A\mu,\quad \delta_B\phi^*_A = \mu
\big(S[\phi,\phi^*]\overleftarrow{\pa}_{\!\!\phi^A}\big),
\quad\delta_B\lambda^A = 0.
\end{eqnarray}
These transformations represent the
BRST transformations in the space of variables $\phi,\;\phi^*,\;\lambda$.
In the case of general gauge theories, the BRST symmetry is not the symmetry of
some action in contrast with the FP method, but as in the case of the Yang-Mills type
of gauge theories, they do not depend on the choice of the gauge-fixing condition.
It is very important to realize that the existence of this symmetry is
the consequence of the fact that the bosonic functional $S$ satisfies the
quantum master equation (\ref{b20}).

The role of this symmetry is
the same as in the case of the Yang-Mills type of gauge theories, namely,
it is responsible for the gauge independence of vacuum functional (\ref{b18}).
Indeed, suppose
$Z_{\Psi}\equiv Z$. We shall change infinitesimally  the gauge
$\Psi\rightarrow\Psi + \delta\Psi$. In the functional integral for
$Z_{\Psi+\delta\Psi}$,
\beq
\label{b25}
Z_{\Psi+ \delta\Psi}=
\int D\phi \; D\phi^*\; d\lambda\; \exp\Big\{\frac{i}{\hbar}\big(S[\phi,\phi^*]
+(\phi^*_A -\Psi[\phi]\overleftarrow{\pa}_{\!\!\!\phi^A})\lambda^A
-
\delta\Psi[\phi]\overleftarrow{\pa}_{\!\!\!\phi^A}\lambda^A\big)\Big\},
\eeq
we make the change of variables in the form of (\ref{b24}) but
with $\mu=\mu[\phi]$ being a functional of $\phi$. The Jacobian
of the transformations in lower order of $\mu[\phi]$ reads
\beq
\label{b26}
J=\exp\big\{-\mu[\phi]\overleftarrow{\pa}_{\!\!\phi^A}\lambda^A+
\mu[\phi]\Delta S[\phi,\phi^*]\big\}.
\eeq
Then, we have
\beq
\nonumber
Z_{\Psi+ \delta\Psi}&=&
\int D\phi \; D\phi^*\; d\lambda\; J\;
\exp\Big\{\frac{i}{\hbar}\big(S[\phi,\phi^*]
+(\phi^*_A -\Psi[\phi]\overleftarrow{\pa}_{\!\!\!\phi^A})\lambda^A
-\\
&&\qquad\qquad\qquad\qquad\quad-
\delta\Psi[\phi]\overleftarrow{\pa}_{\!\!\!\phi^A}\lambda^A
+\mu[\phi]\frac{1}{2}(S,S)\big)\Big\},
\label{b27}
\eeq
Choosing the functional $\mu[\phi]$ in the form
\begin{eqnarray}
\label{b28}
\mu[\phi] = -\frac{i}{\hbar}\delta\Psi[\phi]
\end{eqnarray}
and taking into account that $S[\phi,\phi^*]$ satisfies
the quantum master equation (\ref{b19}), we obtain
\begin{eqnarray}
\label{b29}
Z_{\Psi + \delta\Psi} = Z_{\Psi}.
\end{eqnarray}
In turn, the gauge independence of vacuum functional (\ref{b29}) leads to
the statement about the gauge independence of the S-matrix due to
 the equivalence theorem \cite{KT}. Let us stress once more that the gauge
independence of the vacuum functional (and S-matrix) is a direct consequence of the
BRST symmetry.

\subsection{BRST in FRG}
\noindent
The recent development of quantum field theory is greatly
related with attempts to study nonperturbative aspects of gauge
theories. The request for such a nonperturbative treatment is
related to nonperturbative nature of low-energy QCD and also an
expectation to achieve a consistent theory of quantum gravity. One
of the most promising approaches is related to different versions of
the Wilson renormalization group approach \cite{Wilson,Polch}.
The qualitative idea of this work can be formulated as follows:
regardless, we do not know
how to sum up the perturbative series, in some sense, there is a good
qualitative understanding of the final output of such a summation
for the propagator of the quantum field. A regularized  propagator is
supposed to have a singe pole and also provide some smooth behavior
in the infrared (ir) region. It is possible to write a cutoff-dependent propagator
which satisfies these requirements. Then, the cutoff dependence of
the vertices can be established from the general scale dependence of
the theory, which can be established by means of the functional
methods.
A compact and elegant formulation of the nonperturbative
renormalization group has been proposed  in Refs.
\cite{Wet1,Wet2} in terms of effective average action. The method
was called the FRG approach  for the effective average action; it is nowadays
one of the most popular and developed methods, which can be seen
from the review papers on the FRG approach
\cite{FRG2,FRG1,FRG3,FRG4,IIS-2009,FRG5,FRG6}.

Starting point of the FRG is the action
\beq
\label{c1}
S_{Wk}[\phi]=S_{FP}[\phi]+S_k[\phi],
\eeq
where regulator action $S_k[\phi]$ is constructed by the rule
\beq
\label{c2}
S_k[\phi]=\frac{1}{2}A^iR^{(1)}_{k|ij}A^j+
{\bar C}^{\alpha}R^{(2)}_{k|\alpha\beta}C^{\beta}, \qquad
R^{(1)}_{k|ij}=R^{(1)}_{k|ji}(-1)^{\varepsilon_i\varepsilon_j}.
\eeq
In turn, regulator functions $R^{(1)}_{k|ij}$ and
$R^{(2)}_{k|\alpha\beta}$ obey the properties
\beq
\label{c3}
\lim_{k\rightarrow 0}R^{(1)}_{k|ij}=0,\quad
\lim_{k\rightarrow 0}R^{(2)}_{k|\alpha\beta}=0 \quad
\varepsilon(R^{(1)}_{k|ij})=\varepsilon_i+\varepsilon_j,\quad
\varepsilon(R^{(2)}_{k|\alpha\beta})=
\varepsilon_{\alpha}+\varepsilon_{\beta}.
\eeq
It means that at vanishing regulators the action $S_{Wk}$ coincides
with the FP action,
\beq
\label{c4}
\lim_{k\rightarrow 0}S_{Wk}[\phi]=S_{FP}[\phi].
\eeq
The vacuum functional in the FRG approach is defined with the help of action
$S_{Wk}[\phi]$ in the form of functional integral
\beq
\label{c5}
Z_k=\int D\phi \;\exp\Big\{\frac{i}{\hbar}S_{Wk}[\phi]\Big\}.
\eeq
By construction, the following relation exists
\beq
\label{c6}
\lim_{k\rightarrow 0}Z_{k}[\phi]=Z,
\eeq
where $Z$ is the well-defined vacuum functional in the FP-method
for any Yang-Mills type of gauge theories. The action $S_{Wk}[\phi]$ is not
invariant under the BRST transformations,
\beq
\label{c7}
\delta_B S_{Wk}[\phi]=\delta_B S_k[\phi]\neq 0,
\eeq
where
\beq
\label{c8}
\delta_B S_k[\phi]=
\big(A^iR^{(1)}_{k|ij}R^j_{\alpha}(A)C^{\alpha}-
B^{\alpha}R^{(2)}_{k|\alpha\beta}C^{\beta} -
\frac{1}{2}{\bar C}^{\alpha}R^{(2)}_{k|\alpha\beta}
F^{\beta}_{\gamma\sigma}C^{\sigma}C^{\gamma}
(-1)^{\varepsilon_{\gamma}}\big)\mu .
\eeq
Violation of the BRST symmetry leads to the gauge-dependence problem at least when
$k\neq 0$. Indeed, let $Z_k=Z_{k|\Psi}$ be vacuum functional (\ref{c5})
corresponding to a choice of gauge fixing $\Psi=\Psi[\phi]$.
Consider the vacuum
functional when  the gauge condition is described by functional $\Psi+\delta\Psi$,
\beq
\label{c9}
Z_{k|\Psi+\delta\Psi}=
\int D\phi \;\exp\Big\{\frac{i}{\hbar}\big(S_{Wk}[\phi]+
\delta\Psi[\phi]{\hat R}(\phi)\big)\Big\},
\eeq

Making use of the change of integration variables in the form of BRST
transformation with $\mu[\phi]$ being as in (\ref{b13}), we obtain
\beq
\label{c10}
Z_{k|\Psi+\delta\Psi}= \int D\phi
\;\exp\Big\{\frac{i}{\hbar}\big(S_{Wk}[\phi]+
\delta_BS_k[\phi]\big)\Big\}.
\eeq
We cannot propose a change of
integration variables in (\ref{c10}) to reduce it to $Z_{k|\Psi}$
(see, for example, recent efforts to find a solution
of the problem in gravity theories \cite{BLRNSh}).
So,
\beq
\label{c11} Z_{k|\Psi+\delta\Psi}\neq Z_{k|\Psi}.
\eeq
Therefore, in any case, the gauge-dependence problem exists within
the FRG at the level when $k\neq 0$, and the corresponding S-matrix
does depend on gauges. Violation of the BRST symmetry entails an
additional problem associated with unitarity since the usual
solution assumes the existence of a nilpotent BRST charge \cite{KO}.
Later on, we will return to discussion of this problem when studying
the gauge dependence of  effective  average action.

\section{Ward identities}
\noindent
Quantization of gauge theories leads to very important
understanding concerning the existence of relations between some Green's
functions. These relations in the case of Yang-Mills theories are
known as the Slavnov-Taylor identities \cite{S,T}; for general gauge
theories, they are named as the Ward identities in honor of John Ward
who first discovered an identity in quantum electrodynamics
providing the gradient invariance of the S-matrix elements \cite{Ward}.
In the FRG approach, the relations are refereed as the modified
Slavnov-Taylor identities \cite{Ell}.
Notice that the ST identities
are direct consequence of the gauge invariance of the Yang-Mills action, and
they were introduced before discovery of the BRST symmetry. In
turn, the BRST symmetry helps to present the ST
identities in a unique and compact form (see, for example, Ref.
\cite{LSh} in which this issue is presented and discussed in
details). The latter circumstance is often the cause of
misconception regarding the role of BRST symmetry in the existence
of ST identities.
Our interest in this issue is caused by the widespread opinion
among the FRG community that these identities solve the problem of gauge
dependence. Our point of view is completely different from this
opinion. These identities are direct consequence of the gauge
invariance of the initial classical action on the quantum level providing a
correct solution to the renormalization procedure. Possible misunderstandings
are caused by the fact that these identities can be represented in a universal
form using the BRST transformations. But one must keep in mind
that only in the case when the BRST transformations are
transformations of global supersymmetry of a given gauge
system the gauge independence
of the S-matrix can be confirmed.
In particular, in the case of FRG
approach the mST identities do not guarantee the
BRST symmetry.

\subsection{ST identities in FP method}
\noindent
We begin our discussion of the ST identities appearing as
a direct consequence of gauge invariance of initial classical action
$S_0[A]$.
For all practical goals of quantum calculations in
the case of Yang-Mills type of gauge theories, it is sufficient to
introduce the generating functional of Green's functions
\beq
\label{d1}
Z[j]=\int D\phi
\;\exp\Big\{\frac{i}{\hbar}\big(S_{FP}[\phi]+jA\big)\Big\},
\eeq
where $j_i$, $\varepsilon(j_i)=\varepsilon_i$ are external sources
to fields $A^i$.
Thanks to the gauge invariance of the action $S_0[A]$ (\ref{a1}),
the
Green's functions of the theory obey the relations known as the
ST identities \cite{S,T}. These identities can be
derived  from (\ref{d1}) by means of the change of integration
variables $A^i$, in the form of infinitesimal gauge
transformations (\ref{a1}). The Jacobian of these transformations is
equal to unity. Then, the basic ST identities for
Yang-Mills fields can be written in the form
\beq
\nonumber
&&j_i \langle R^i_{\alpha} (A)\rangle_j \,+\, \langle
B^{\beta}\big(\chi_{\alpha}(A,B)
\overleftarrow{\pa}_{\!\!A^i}\big)R^i_{\alpha}(A)\rangle_j \,+\\
\nonumber
&&+
\langle {\bar C}^{\beta}\big(\chi_{\beta}(A,B)
\overleftarrow{\pa}_{\!\!A^i}\big)
R^i_{\gamma,k}(A)R^k_{\alpha}(A)C^{\gamma}\rangle_j
(-1)^{\varepsilon_{\alpha}(\varepsilon_{\gamma}+1)}
\, -\\
\label{d2}
&&
-\langle {\bar
C}^{\beta}\big(\chi_{\beta}(A,B)
\overleftarrow{\pa}_{\!\!A^i}\overleftarrow{\pa}_{\!\!A^k}\big)
R^k_{\gamma}(A)C^{\gamma}R^i_{\gamma}(A) \rangle_j
(-1)^{\varepsilon_i+\varepsilon_j}
\,\equiv \,0\,,
\eeq
where the symbol \ $\langle G(\phi) \rangle_j$ \
means the vacuum expectation value of
the quantity \ $G(\phi)$ \ in the presence of external sources \
$j^a_{\mu}$,
\beq
\label{d3}
\langle G(\phi) \rangle_j \,=\, \int
D\phi\,\, G(\phi) \, \exp \Big\{
\frac{i}{\hbar}\big[S_{FP}[{\phi}]+jA\big]\Big\}\,.
\eeq
The
generating functional \ $Z[j]$ \ contains information about all
Green's functions of the theory, which can be obtained by taking
variational derivatives with respect to the sources. Similarly, the
ST identities represent an infinite set of relations
obtained from (\ref{d2}) by taking derivatives with respect to
external sources \ $j^a_{\mu}$. In the case of linear gauge condition, the last
summand in (\ref{d2}) disappears.

The form of the ST identities can be greatly simplified
by introducing extra sources to the ghost, antighost, and auxiliary
fields. In this case, one has to deal with the extended generating
functional of the theory
\beq
\label{d4}
Z[J]\,=\, \int D{\phi} \;{\rm exp}
\Big\{\frac{i}{\hbar}\big[S_{FP}[{\phi}]+J\phi\big]\Big\}\,.
\eeq
The generating functional of connected Green's functions,
$W[J]$, is defined by the relation
\beq
\label{d5}
Z[J]={\rm exp} \Big\{
\frac{i}{\hbar}\,W[J] \Big\}\,.
\eeq
Finally, the generating functional of the vertex Green's functions
(effective action) is defined through the Legendre transformation of
 $W[J]$,
\beq
\label{d6}
\Gamma[\Phi]=W[J] -J\Phi,
\eeq
where the source
fields  $J_A$  are solutions of the equations
\beq
\label{d7}
\Phi^A=\overrightarrow{\pa}_{\!\!J_A}W[J].
\eeq
By means of
(\ref{d6}) and (\ref{d7}), one can easily arrive at the relations
\beq
\label{d8}
\Gamma[\Phi]\overleftarrow{\pa}_{\!\!\Phi^A}=-J_A.
\eeq
The ST identities which are consequences of gauge
symmetry of initial action can be rewritten with the help of the BRST
symmetry of the Faddeev-Popov action. For this end, we make use of
the change of variables in the functional integral (\ref{d4}) of the
form (\ref{b6}). Because of the  property (\ref{b16})
 and nilpotency of  $\mu$,
the Jacobian of this transformation is equal to 1. Using the
invariance of the functional integral under change of integration
variables, the following identity holds
\beq
\label{d9}
\int
D\phi \,J\delta_B\phi \; \exp \Big\{ \frac{i}{\hbar}\,\big(S_{FP}[\phi] +
J\phi\big) \Big\}
 \,\equiv\, 0\,.
\eeq
Here, the nilpotency of BRST transformation and the consequent
exact relation
\beq
\label{d10}
\exp \Big\{ \frac{i}{\hbar}J\delta_B\phi
\Big\}= 1 + \frac{i}{\hbar} J\delta_B\phi
\eeq
have been used.

From (\ref{d5}) and (\ref{d8}), it follows
\beq
\label{d11}
J_AR^A\big(-i\hbar\overrightarrow{\pa}_{\!\!J}\big)Z[J]\equiv 0, \qquad
J_AR^A\big(-i\hbar\overrightarrow{\pa}_{\!\!J}\big)W[J]\equiv 0,
\eeq
which are the ST identities in a closed form for the functionals $Z[J]$ and $W[J]$.
These identities, like those (\ref{d2}), contain explicit information about
gauge theory  through  generators of the BRST transformations. There exists
a possibility
to present the ST identities in a unique form with  the introduction of
a set of external sources (known as antifields in the BV formalism)
$\Phi^*_A,\quad\vp(\Phi^*_A)=\vp_A+1$
to the BRST transformations and the extended generating functional of
Green's functions
\beq
\label{d13}
Z[J,\Phi^*]=\int D{\phi}\;
\exp \Big\{ \frac{i}{\hbar}\,\big[S_{FP}[\phi] + J\phi
+ \Phi^*_A R^A(\phi)\big]\Big\}=\exp\Big\{ \frac{i}{\hbar}W[J,\Phi^*]\Big\},,
\eeq
where we used the notation for BRST transformations, $R^A(\phi)$, which
was previously introduced in (\ref{b6}). It is clear that
\beq
\label{d14}
Z[J,\Phi^*]\Big|_{\Phi^*=0} = Z[J].
\eeq
Now, we can present the ST identities (\ref{d11}) in the following form
\beq
\label{d15}
J_A\overrightarrow{\pa}_{\!\!\Phi^*_A}Z[J,\Phi^*]\equiv\, 0,\qquad
J_A\overrightarrow{\pa}_{\!\!\Phi^*_A}W[J,\Phi^*]\equiv\, 0,.
\eeq
In terms of the extended effective action, $\Gamma=\Gamma[\Phi,\Phi^*]$,
\beq
\label{d16}
\Gamma[\Phi,\Phi^*]=W[J,\Phi^*]-J\Phi,\quad
\Phi^A=\overrightarrow{\pa}_{\!\!J_A}W[J,\Phi^*], \quad
\Gamma[\Phi,\Phi^*]\overleftarrow{\pa}_{\!\!\Phi^A}=-J_A,
\eeq
the identities (\ref{d15}) is rewritten as
\beq
\label{d17}
\Gamma\overleftarrow{\pa}_{\!\!\Phi^A}
\overrightarrow{\pa}_{\!\!\Phi^*_A}\Gamma\equiv\, 0
\eeq
in the form of a nonlinear equation with respect to $\Gamma$
(in the form of the Zinn-Justin equation \cite{Z-J}).

\subsection{Ward identities in BV formalism}
\noindent
Now, we shall proceed with the derivation of the Ward identity
for general gauge theories within the BV formalism.
It is very useful from the beginning to work with
 the extended generating functional of Green's
functions
\begin{eqnarray}
\label{d18}
Z[J,\phi^*]=
\int
D{\phi}\exp\Big\{\frac{i}{\hbar}\big(S_{ext}[\phi,\phi^*]+J_A\phi^A\big)\Big\}=
\exp\Big\{\frac{i}{\hbar}W[J,\phi^*]\Big\},
\end{eqnarray}
where $W[J,\phi^*]$ is the generating functional for connected Green's functions,
\begin{eqnarray}
\label{d19}
S_{ext}[\phi, \phi^*]= S[\phi,\phi^*+
\Psi[\phi]\overleftarrow{\pa}_{\!\!\phi}],
\end{eqnarray}
and functional $S[\phi,\phi^*]$ satisfies the quantum master-equation
(\ref{b18}) and the boundary condition (\ref{b19}).
Gauge-fixing procedure (\ref{d18}) used in the BV formalism \cite{BV,BV1}
can be described in terms of anticanonical transformation,
\beq
\label{d20}
\phi^{'A}=\overrightarrow{\pa}_{\phi^{*'}_A}F[\phi,\phi^{*'}],\quad
\phi^*_A=F[\phi,\phi^{*'}]\overleftarrow{\pa}_{\phi^A},
\eeq
of a special form corresponding to the choice of generating functional
$F[\phi,\phi^{*'}]$ in the form,
\beq
\label{d21}
F[\phi,\phi^{*'}]=\phi^{*'}_A\phi^A+\Psi[\phi], \quad \vp(\Psi)=1,
\eeq
as it was proposed for the first time in Ref. \cite{VLT}.

Notice that the action $S_{\rm ext}[\phi,\phi^*]$
satisfies  the quantum master equation (\ref{b18}) as well. Indeed, the equality holds,
\footnote{For any two quantities $F$ and $H$, the  supercommutator is defined as
$[F,H]=FH-HF(-1)^{\varepsilon(F)\varepsilon(H)}$.}
\begin{eqnarray}
\label{d22}
\exp\Big\{\frac{i}{\hbar}S_{\rm
ext}[\phi,\;\phi^*]\Big\}= \exp\{[\Psi,\;\Delta]\}
\exp\Big\{\frac{i}{\hbar}S[\phi,\;{\phi}^*]\Big\},
\end{eqnarray}
because
\begin{eqnarray}
\label{d23}
[\Psi,\;\Delta]= \Psi\overleftarrow{\pa}_{\!\!{\phi}^A}
\overrightarrow{\pa}_{\!\!{\phi}^*_A},
\end{eqnarray}
and the operator $\exp\{[\Psi,\;\Delta]\}$ acts as the translation
operator with respect to ${\phi}^*_A$. Note that
\begin{eqnarray}
\label{d24}
[\Delta,\;[\Psi,\;\Delta]] = 0,
\end{eqnarray}
and therefore
\begin{eqnarray}
\label{d25}
\Delta \exp\Big\{\frac{i}{\hbar}S_{\rm ext}\Big\}=0\quad\rightarrow \quad
\frac{1}{2}(S_{\rm ext},S_{\rm ext})=i\hbar \Delta S_{\rm ext}.
\end{eqnarray}

Taking into account the equation  (\ref{d25}), the explicit form
of the operator $\Delta$ (\ref{b22})  and independence of operator
$\overrightarrow{\pa}_{\!\!\phi^*_A}$
on the integration variables in functional integral we have  the evident relation
\begin{eqnarray}
\nonumber
0&=&
\int D{\phi}\exp\Big\{\frac{i}{\hbar}J_A\phi^A\Big\}
\Delta\exp\Big\{\frac{i}{\hbar}S_{\rm ext}[\phi,\phi^*]\Big\}\\
\label{d26}
&=&(-1)^{\varepsilon_A}\overrightarrow{\pa}_{\!\!\phi^*_A}
\int D{\phi}\exp\Big\{\frac{i}{\hbar}J_A\phi^A\Big\}
\overrightarrow{\pa}_{\!\!\phi^A}\exp\Big\{\frac{i}{\hbar}
S_{\rm ext}[\phi,\phi^*]\Big\}.
\end{eqnarray}
Integrating by parts in the last integral, one finds that the theory
in question satisfies the equality
\begin{eqnarray}
\label{d27}
J_A \overrightarrow{\pa}_{\!\!\phi^*_A}Z[J,\phi^*] = 0.
\end{eqnarray}
This is the  Ward identity written for the extended generating
functional of Green's functions.
For the generating functional of connected Green's functions
$W[J, \phi^*]$,
the identity (\ref{d27}) is rewritten in the form
\begin{eqnarray}
\label{d28}
J_A \overrightarrow{\pa}_{\!\!\phi^*_A}W[J,\phi^*] = 0.
\end{eqnarray}
Introducing the generating functional of the
vertex functions $\Gamma = \Gamma [\Phi,\;\Phi^*]$
(for uniformity of notations, we
use ${\phi}^*_A={\Phi}^*_A$) in a standard manner,
through the Legendre transformation
of $W[J,\Phi^*]$,
\begin{eqnarray}
\label{d29}
\Gamma [\Phi,\;\Phi^*] = W[J,\Phi^*] - J_A\Phi^A,\quad
\Phi^A = \overrightarrow{\pa}_{\!\! J_A}W[J,\Phi^*],
\quad \Gamma[\Phi,\Phi^*]\overleftarrow{\pa}_{\!\!\phi^A}=-J_A.
\end{eqnarray}
 the Ward identity (\ref{d28}) for $\Gamma =\Gamma [\Phi,\;\Phi^*]$
 takes the form
 of classical master equation,
\begin{eqnarray}
\label{d30}
(\Gamma,\Gamma) = 0.
\end{eqnarray}
The form (\ref{d30}) coincides with (\ref{d17}). The Ward identity (\ref{d30}) plays
a crucial role in proving the gauge invariant renormalizability of general gauge theories
\cite{VLT}.

\subsection{Modified Slavnov-Taylor  identities in FRG}
\noindent
Although the BRST symmetry is broken in the FRG approach, nevertheless,
certain relations between the Green's functions known as the mST identities exist.
It confirms that the existence of these relations is not related
with  the BRST symmetry but the main reason is gauge invariance
of an initial classical action.

To discuss the mST identities, it is useful
as in previous cases to introduce the average generating functional
of Green's functions $Z_k=Z_k[J,\Phi^*]$ and the
average generating functional of connected Green functions $W_k=W_k[J,\Phi^*]$
in the FRG approach,
\beq
\nonumber
Z_k[J,\Phi^*]&=&\int D\phi
\exp\Big\{\frac{i}{\hbar}\big(S_0[A]+S_k[\phi]+
\Psi[\phi]{\hat R}(\phi)
+J_A\phi^A+\Phi^*_AR^A(\phi)\big)\Big\}=\\
\label{d31}
&=&
\exp\Big\{\frac{i}{\hbar}W_k[J,\Phi^*]\Big\}.
\eeq
Making use of the change of integration variables in the sector of fields $A^i$
in the form of gauge transformations
\beq
\label{d32}
\delta A^i=R^i_{\alpha}(A)C^{\alpha}\mu=R^i(\phi)\mu,
\eeq
taking into account the invariance of $S_0[A]$ under transformations (\ref{d32})
and the Jacobian of these transformations
\beq
\label{d33}
J=1+(-1)^{\varepsilon_i}\overrightarrow{\pa}_{\!\!A^i}
R^i_{\alpha}(A)C^{\alpha}\mu,
\eeq
we arrive at the identity
\beq
\nonumber
&&\big(J_j\overrightarrow{\pa}_{\!\!\Phi^*_j}+
S_{k,j}[-i\hbar\overrightarrow{\pa}_{\!\!J}]
\overrightarrow{\pa}_{\!\!\Phi^*_j}+
(-1)^{\varepsilon_j(\varepsilon_{\alpha}+1)}
R^j_{\alpha,j}(-i\hbar\overrightarrow{\pa}_{\!\!J})
\overrightarrow{\pa}_{\!\!{\bar J}_{\alpha}}+
\Phi^*_AR^A_{,j}(-i\hbar\overrightarrow{\pa}_{\!\!J})
\overrightarrow{\pa}_{\!\!\Phi^*_j}+\\
&&+ \Psi_{,A}[-i\hbar\overrightarrow{\pa}_{\!\!J}]
R^A_{,i}(-i\hbar\overrightarrow{\pa}_{\!\!J})
\overrightarrow{\pa}_{\!\!\Phi^*_j}+
(-1)^{\varepsilon_j}\Psi_{,jA}[-i\hbar\overrightarrow{\pa}_{\!\!J}]
\overrightarrow{\pa}_{\!\!\Phi^*_A}
\overrightarrow{\pa}_{\!\!\Phi^*_j} \big)Z_k[J,\Phi^*]\equiv 0,
\label{d34}
\eeq
which is nothing but the mST identity in the FRG
approach and a direct consequence of gauge invariance of initial
classical action $S_0[A]$ at the quantum level. Note that the mST
identity in the case of pure Yang-Mills theory formulated in linear
nonsingular Lorenz invariant gauges for the FRG approach was
derived in  \cite{Ell}.

One can present the mST identity (\ref{d34}) in a more compact form
using additional information about invariance properties of quantities entering
the exponent of the integrand (\ref{d31}). Consider the change of variables
$C^{\alpha}$, ${\bar C}^{\alpha}$,
\beq
\label{d35}
\delta C^{\alpha}=-\frac{1}{2}(-1)^{\varepsilon_{\beta}}
F^{\alpha}_{\beta\gamma}
C^{\gamma }C^{\beta}\mu, \quad
\delta {\bar C}^{\alpha}=\mu B^{\alpha}
\eeq
 in the functional integral entering the identity (\ref{d34}). Then, the result
 \beq
 \label{d36}
\big(J_A\overrightarrow{\pa}_{\!\!\Phi^*_A}+
S_{k,A}[-i\hbar\overrightarrow{\pa}_{\!\!J}]
\overrightarrow{\pa}_{\!\!\Phi^*_A}
\big)Z_k[J,\Phi^*]\equiv 0
 \eeq
coincides with that obtained by making use the change of variables $\phi^A$ in the form
of the BRST transformations, $\delta\phi^A=R^A(\phi)\mu$
in the functional (\ref{d31}).
In terms of the average generating functional of connected Green's functions,
$W_k=W_k[J,\Phi^*]$, the mST identity
(\ref{d36}) is rewritten as
\beq
 \label{d38}
\big(J_A\overrightarrow{\pa}_{\!\!\Phi^*_A}+
S_{k,A}[(\overrightarrow{\pa}_{\!\!J}W_k)-i\hbar\overrightarrow{\pa}_{\!\!J}]
\overrightarrow{\pa}_{\!\!\Phi^*_A}
\big)W_k[J,\Phi^*]\equiv 0 .
 \eeq
The  effective average action, $\Gamma_k=\Gamma_k[\Phi,\Phi^*]$,
is defined through the
Legendre transformation of $W_k$,
\beq
 \label{d39}
 \Gamma_k[\Phi,\Phi^*]=W_k[J,\Phi^*]-J\Phi,\quad
 \Phi^A=\overrightarrow{\pa}_{\!\!J_A}W_k[J,\Phi^*],\quad
\Gamma_k[\Phi,\Phi^*]\overleftarrow{\pa}_{\!\!\Phi^A}=-J_A .
\eeq
Then, the mST identity (\ref{d38}) can be presented in terms of $\Gamma_k$ as
\beq
\label{d40}
\Gamma_k\overleftarrow{\pa}_{\!\!\Phi^A}\overrightarrow{\pa}_{\!\!\Phi^*_A}\Gamma_k-
S_{k,A}[{\hat \Phi}]
\overrightarrow{\pa}_{\!\!\Phi^*_A}\Gamma_k\equiv 0 ,
\eeq
or, using the antibracket,
\beq
\label{d41}
\frac{1}{2}(\Gamma_k,\Gamma_k)-
S_{k,A}[{\hat \Phi}]
\overrightarrow{\pa}_{\!\!\Phi^*_A}\Gamma_k\equiv 0 ,
\eeq
where the notations
\beq
\label{d42}
{\hat \Phi}^A=\Phi^A+ i\hbar(\Gamma^{''-1}_k)^{AB}\,
\overrightarrow{\pa}_{\!\!\Phi^B},\quad
(\Gamma_k^{''})_{AB}=\overrightarrow{\pa}_{\!\!\Phi^A}\Gamma_k
\overleftarrow{\pa}_{\!\!\Phi^B},\quad
\big(\Gamma^{''-1}_k\big)^{AC}\cdot
\big(\Gamma^{''}_k\big)_{CB}\,=\delta^A_{\,B},
\eeq
are used. In the limit $k\rightarrow 0$, the mST identity (\ref{d41}) reduces to (\ref{d30}).

\section{Gauge dependence}
\noindent
The gauge dependence is a problem in quantum description
of gauge theories. Any covariant quantization scheme
(FP method \cite{FP}, BV formalism
\cite{BV,BV1}, FRG approach \cite{Wet1,Wet2},
Gribov-Zwanziger theory \cite{Gribov,Zwanziger,Zwanziger1})
for gauge theories meets with the
gauge-dependence problem. Here, we remember the main
aspects and solutions of  the gauge-dependence problem in the
FP method and the BV formalism. We obtain new results concerning the
gauge-dependence problem of  the effective average action precisely on
the  level of the flow equation.

\subsection{Gauge dependence in FP method}
\noindent
It is well known that Green's functions in gauge theories depend on
the choice of gauge \cite{J,DeW,Niel,GvNW,FK,Boul,LT3,LT1,TY,LR,BG}.
From the gauge independence of the $S$-matrix (see (\ref{b15})), it
follows that the gauge dependence of Green's functions in gauge
theories must be of a special character. To study the character of
this dependence, let us consider an infinitesimal variation of gauge-fixing
functional $\Psi[\phi]\;\rightarrow\;\Psi[\phi]+\delta\Psi[\phi]$ in
the functional integral (\ref{b12}). Then, we obtain
\beq
\label{d43}
\delta Z[J,\Phi^*]=\frac{i}{\hbar}\int D\phi\; \delta\Psi_{,A}[\phi]R^A(\phi)
\;\exp\Big\{\frac{i}{\hbar}\big(S_{PF}[\phi]
+J_A\phi^A+\Phi^*_AR^A(\phi)\big)\Big\}.
\eeq
Making use of the change of integration variables in the functional integral
(\ref{d43}) in the form of the BRST transformations,
\beq
\label{d44}
\delta\phi^A=R^A(\phi)\mu[\phi],
\eeq
taking into account that due to (\ref{b17}) the corresponding Jacobian, $J$,
is equal to
\beq
\label{d45}
J=\exp\{-\mu[\phi]_{,A}R^A(\phi)\},
\eeq
choosing the functional $\mu[\phi]$ in the form
$\mu[\phi]=(i/\hbar)\delta\Psi[\phi]$, the relation (\ref{d43}) is rewritten as
\beq
\nonumber
\delta Z[J,\Phi^*]&=&\frac{i}{\hbar}\int D\phi\;
J_AR^A(\phi)\delta\Psi[\phi]\exp\Big\{\frac{i}{\hbar}\big(S_{PF}[\phi]
+J_A\phi^A+\Phi^*_AR^A(\phi)\big)\Big\}=\\
\label{d46}
&=&
\frac{i}{\hbar}J_AR^A(-i\hbar\overrightarrow{\pa}_{\!\!J})\;
\delta\Psi[-i\hbar\overrightarrow{\pa}_{\!\!J}]\;Z[J,\Phi^*].
\eeq
The Eq. (\ref{d43}) can be equivalently presented in the form
\beq
\label{d47}
\delta Z[J,\Phi^*]=\frac{i}{\hbar}
\delta\Psi_{,A}[-i\hbar\overrightarrow{\pa}_{\!\!J}]
R^A(-i\hbar\overrightarrow{\pa}_{\!\!J})Z[J,\Phi^*].
\eeq
The relations (\ref{d46}) and (\ref{d47}) are equivalent
due to the evident equality
\beq
\label{d48}
\int D\phi \overrightarrow{\pa}_{\!\!\phi^B}\Big(\Psi[\phi]R^B(\phi)
\exp\Big\{\frac{i}{\hbar}\big(S_{PF}[\phi]
+J_A\phi^A+\Phi^*_AR^A(\phi)\big)\Big\}\Big)=0,
\eeq
where the equations
\beq
\label{d49}
S_{PF,A}[\phi]R^A(\phi)=0,\quad R^A_{\;,A}(\phi)=0,\quad
R^A_{\;,B}(\phi)R^B(\phi)=0,
\eeq
should be used.
In terms of the functional $W[J,\Phi^*]$, the relations
(\ref{d46}) and (\ref{d47}) are rewritten as
\beq
\label{d50}
\delta W[J,\Phi^*]=J_A
R^A(\overrightarrow{\pa}_{\!\!J}W-i\hbar\overrightarrow{\pa}_{\!\!J})\;
\delta\Psi[\overrightarrow{\pa}_{\!\!J}W-i\hbar\overrightarrow{\pa}_{\!\!J}]
\cdot 1,
\eeq
and
\beq
\label{d51}
\delta W[J,\Phi^*]=
\delta\Psi_{,A}[\overrightarrow{\pa}_{\!\!J}W-i\hbar\overrightarrow{\pa}_{\!\!J}]
R^A(\overrightarrow{\pa}_{\!\!J}W-i\hbar\overrightarrow{\pa}_{\!\!J})\cdot 1.
\eeq
Finally, the gauge dependence of the effective action,
$\Gamma=\Gamma[\Phi,\Phi^*]$, is described by the relation
\beq
\label{d52}
\delta\Gamma[\Phi,\Phi^*]=-(\Gamma\overleftarrow{\pa}_{\!\!\Phi^A})\;
R^A({\hat \Phi})\;\delta\Psi[{\hat \Phi}]\cdot 1,
\eeq
or
\beq
\label{d53}
\delta\Gamma[\Phi,\Phi^*]=
\delta\Psi_{,A}[{\hat \Phi}]\;
R^A({\hat \Phi})\cdot 1.
\eeq
Calculating the effective action $\Gamma[\Phi,\Phi^*]$
on its extremals $\pa_{\Phi^A}\Gamma=0$,
from the equation (\ref{d52}) it follows that this action
does not depend on the gauges,
\beq
\label{d54}
\delta\Gamma\big|_{\pa_{\Phi}\Gamma=0}=0,
\eeq
making possible the physical interpretation of results obtained in the
FP method.

\subsection{Gauge dependence in BV formalism}
\noindent
Let us consider the gauge dependence problem in the BV-formalism.
To do this we make an infinitesimal variation of the
gauge fixing functional $\Psi[\phi]\rightarrow\Psi[\phi] +
\delta\Psi[\phi]$. Then due to (\ref{d22}), the variation of
$\exp\{(i/\hbar)S_{\rm
ext}\}$ reads
\begin{eqnarray}
\label{e1}
\delta\Big(\exp\Big\{\frac{i}{\hbar}S_{\rm ext}\Big\}\Big) =
[\delta\Psi,\;\Delta]\;
\exp\Big\{\frac{i}{\hbar}S_{\rm ext}\Big\} = \Delta\;\delta\Psi
\;\exp\Big\{\frac{i}{\hbar}S_{\rm ext}\Big\}
\end{eqnarray}
because in the case, when $\Psi$ and $\delta\Psi$ depend on the variables
$\phi$ only, the operator $[\delta\Psi,\;\Delta]$ commutes with
$[\Psi,\;\Delta]$.

Next, the corresponding variation of the functional
$Z[J,\;{\Phi}^*]$ has the form
\begin{eqnarray}
\nonumber
\delta Z[J,\;{\phi}^*]& =& \int d\phi\;
\exp\Big\{\frac{i}{\hbar}J_A\phi^A\Big\}
\Delta\;\delta\Psi\; \exp\Big\{\frac{i}{\hbar}
S_{\rm ext}(\phi,\;{\phi}^*)\Big\}= \\
\nonumber &=&
(-1)^{\varepsilon_A}\overrightarrow{\pa}_{\!\!\phi^*_A}
\int d\phi\; \exp\Big\{\frac{i}{\hbar}J_A\phi^A\Big\}
\overrightarrow{\pa}_{\!\!\phi^A}\;\delta\Psi\;
\exp\Big\{\frac{i}{\hbar}S_{\rm ext}(\phi,\;{\phi}^*)\Big\}= \\
\label{e2}
 &=&
- \overrightarrow{\pa}_{\phi^*_A} J_A
\int d\phi\;\delta\Psi\;
\exp\Big\{\frac{i}{\hbar}\Big[S_{\rm ext}(\phi,\;{\phi}^*) +
J_A\phi^A\Big]\Big\}.
\end{eqnarray}
Therefore
\begin{eqnarray}
\label{e3}
\delta Z[J,\phi^*]=
-\frac{i}{\hbar}J_A\overrightarrow{\pa}_{\!\!\phi^*_A}\;
\delta\Psi[-i\hbar\overrightarrow{\pa}_{\!\!J}] Z[J,\;{\phi}^*].
\end{eqnarray}
In terms of the generating functional $W=W[J,\;{\phi}^*]$
of connected Green's functions,
we have
\begin{eqnarray}
\label{e4}
\delta W[J,\phi^*]= -J_A
\overrightarrow{\pa}_{\!\!\phi^*_A}\;
\Psi[(\overrightarrow{\pa}_{\!\!J}W)-i\hbar\overrightarrow{\pa}_{\!\!J}]
\cdot 1.
\end{eqnarray}
In deriving the relation (\ref{e4}) describing the gauge dependence of
functional $W$ the Ward identity (\ref{d15}) has been substantially used.
This once again emphasizes that the gauge dependence problem cannot
be reduced to fulfilling Ward's identities.
The variation of the generating functional of vertex functions
$\Gamma = \Gamma [\Phi,\;\Phi^*]$, where $\Phi^*_A=\phi^*_A$,
$\Phi^A=\overrightarrow{\pa}_{\!\!J_A}W[J,\Phi^*]$, can be written as
\begin{eqnarray}
\label{e5}
\delta\Gamma =\Gamma\overleftarrow{\pa}_{\!\! \Phi^A}\Big(
\overrightarrow{\pa}_{\!\!\Phi^*_A}\langle\delta\Psi\rangle
+(\overrightarrow{\pa}_{\!\!\Phi^*_A} \Phi^B)\overrightarrow{\pa}_{\!\!\Phi^B}
\langle\delta\Psi\rangle\Big),
\end{eqnarray}
where we have used the equality
\begin{eqnarray}
\label{e6}
\overrightarrow{\pa}_{\!\!\Phi^*_A}\Big|_J=
\overrightarrow{\pa}_{\!\!\Phi^*_A}\Big|_{\Phi}
+ (\overrightarrow{\pa}_{\!\!\Phi^*_A}\Phi^B)\Big|_J
\overrightarrow{\pa}_{\!\!\Phi^B}\Big|_{\Phi^*},
\end{eqnarray}
and also introduced the notation $\langle\delta\Psi\rangle=
\langle\delta\Psi\rangle[\Phi,\Phi^*]$ for the functional
\beq
\label{e7}
\langle\delta\Psi\rangle=
\delta\Psi[{\hat \Phi}]\cdot 1, \quad {\hat \Phi}^A=\Phi^A+
i\hbar(\Gamma^{''-1})^{AB}\overrightarrow{\pa}_{\!\!\Phi^B}
\eeq
where
\beq
\label{e8}
\Gamma^{''}_{AB}=\overrightarrow{\pa}_{\!\!\Phi^A}\Gamma
\overleftarrow{\pa}_{\!\!\Phi^B},\quad
\big(\Gamma^{''-1}\big)^{AC}\cdot
\Gamma^{''}_{CB}\,=\delta^A_{\;\;B}.
\eeq

From (\ref{e5}) it follows very important statement that the effective action
$\Gamma[\Phi,\Phi^*]$ does not depend on gauge conditions at their extremals,

Calculating the effective action $\Gamma[\Phi,\Phi^*]$ on its
extremals $\pa_{\Phi^A}\Gamma=0$,
from the equation (\ref{e5}) it follows that this action
does not depend on the gauges
\begin{eqnarray}
\label{e9}
\delta\Gamma[\Phi,\Phi^*]\Big|_{\pa_{\Phi}\Gamma=0} =0.
\end{eqnarray}

There is another point of view related  with this fact. Indeed, taking into
account the Ward identity for the functional $W=W[J,\Phi^*]$ (\ref{d15})
 we derive the relations
\begin{eqnarray}
\label{e10}
0=\overrightarrow{\pa}_{\!\!J_B}\big(J_A\overrightarrow{\pa}_{\!\!\Phi^*_A}W\big)=
\overrightarrow{\pa}_{\!\!\Phi^*_A}W+
(-1)^{\varepsilon_B}J_A\overrightarrow{\pa}_{\!\!\Phi^*_A}
\overrightarrow{\pa}_{\!\!J_B}W,\quad
J_A\overrightarrow{\pa}_{\!\!\Phi^*_A}\Phi^B=J_A\overrightarrow{\pa}_{\!\!\Phi^*_A}
\overrightarrow{\pa}_{\!\!J_B}W.
\end{eqnarray}
Therefore, we can rewrite the equation (\ref{e5}) in the form
\begin{eqnarray}
\label{e11}
\delta\Gamma=\Gamma\big(\overleftarrow{\pa}_{\!\!\Phi^A}\overrightarrow{\pa}_{\!\!\Phi^*_A}-
\overleftarrow{\pa}_{\!\!\Phi^*_A}\overrightarrow{\pa}_{\!\!\Phi^A}\big)
\langle\delta\Psi\rangle=
(\Gamma,\langle\delta\Psi\rangle).
\end{eqnarray}
We  see that the variation of the functional $\Gamma$ under an infinitesimal
change of gauge fixing may be expressed in the form of anticanonical
transformation (\ref{d20}) of the fields and antifields with the
generating
function $F=F(\Phi,\;\Phi^*)=
\Phi^*_A\Phi^A + \langle\delta\widehat\Psi\rangle$
\begin{eqnarray}
\label{e12}
{\Phi^{'A}}= \Phi^A + \overrightarrow{\pa}_{\!\!\Phi^*_A}
\langle\delta{\Psi}\rangle,\quad
{\Phi^{*^{'}}_A}= \Phi^*_A -
\langle\delta{\Psi}\rangle\overleftarrow{\pa}_{\!\!\Phi^A}.
\end{eqnarray}
For the first time such character of gauge dependence of the effective
action in the BV-formalism has been described in \cite{VLT} allowing to prove
gauge invariant renormalizability of general gauge theories.

\subsection{Gauge dependence in FRG}

We consider the gauge dependence problem within the FRG approach not restricting
yourself by  special types of initial classical action, $S_0[A]$,
or gauge fixing condition, $\Psi[\phi]$.
We demonstrate that
derivation of flow equation and analysis of gauge dependence
have the same level of accuracy.

The generating functional of Green functions has the form
\beq
\label{e13}
Z_k[J,\Phi^*]=\int D\phi \,\exp \Big\{\frac{i}{\hbar}
\big[S_{Wk}[\phi]+\Phi^*_AR^A(\phi)
+J_A\phi^A\big]\Big\}=\exp \Big\{\frac{i}{\hbar}
W_{k}[J,\Phi^*]\Big\},
\eeq
where
\beq
\label{e14}
S_{Wk}[\phi]=S_0[A]+S_k[\phi]+\Psi_{,A}[\phi]R^A(\phi).
\eeq
Let us find the partial derivative of $Z_k[J,\Phi^*]$ with respect
to IR cutoff parameter $k$.
The result reads
\beq
\nonumber
\pa_kZ_k[J,\Phi^*]&=&\frac{i}{\hbar}\int D\phi \,
\pa_kS_k[\phi]
\exp \Big\{\frac{i}{\hbar}
\big[S_{Wk}[\phi]+\Phi^*_AR^A(\phi)
+J_A\phi^A\big]\Big\}=\\
\label{e15}
&=&\frac{i}{\hbar}
\pa_kS_k[-i\hbar\overrightarrow{\pa}_{\!\!J}]Z_k[J,\Phi^*].
\eeq
In deriving this result, the existence of functional integral (\ref{e13})
is only used.
In terms
of generating functional of connected Green functions we have
\beq
\label{e16}
\pa_k W_k[J,\Phi^*]=\pa_k S_k[\overrightarrow{\pa}_{\!\!J}W_k
-i\hbar\overrightarrow{\pa}_{\!\!J}]\cdot 1 .
\eeq
The basic equation (flow equation)  of the FRG approach follows from (\ref{e16})
\beq
\label{e17}
\pa_k \Gamma_k[\Phi,\Phi^*]=\pa_k S_k[{\hat \Phi}]\cdot 1,
\eeq
where ${\hat \Phi}=\{{\hat \Phi}^A\}$ is defined in (\ref{d42}). It follows from
(\ref{d42}) that $\pa_k{\hat \Phi}^A\neq 0$. It is assumed that solutions
to the flow
equations (\ref{e17}) present the effective average action
$\Gamma_k[\Phi,\Phi^*]$
beyond the usual perturbation calculations. In perturbation theory
the functional \ $\Gamma_k=\Gamma_k[\Phi,\Phi^*]$ is considered as a
 solution to the functional
integro-differential equation
\beq
\label{e18}
\!\!\exp \Big\{ \frac{i}{\hbar}\,\Ga_k[\Phi,\Phi^*] \Big\}
\!\!=\!\!
\int D\phi \,\exp \Big\{\frac{i}{\hbar}
\big[S_{Wk}[{\Phi+\phi}]+\Phi^*_AR^A(\Phi+\phi)
-\Ga_k[\Phi,\Phi^*]\overleftarrow{\pa}_{\!\!\Phi^A}\,\phi^A\big]\Big\},
\eeq
using  in the functional integral  the Taylor expansion for the exponent
with respect to fields $\phi$, and then integrating  over  $\phi$.
Such procedure is mathematical correct because the functional integral
is well defined in the perturbation theory \cite{Slavnov}. It is known fact
\cite{LSh} that the  effective average action found as a solution to the equation
(\ref{e18}) depends on gauges even on-shell.

Now, we analyze the gauge dependence problem of
the flow equation (\ref{e17}). Note that up to now this problem has never been discussed
in the literature.
To do this we consider the variation of $\pa_kZ_k[J,\Phi^*]$ (\ref{e15})
under an infinitesimal change
of gauge fixing functional, $\Psi[\phi]\rightarrow \Psi[\phi]+\delta\Psi[\phi]$.
Taking into account that $\pa_kS_k$ does not depend on gauge fixing procedure, we obtain
\beq
\label{e19}
\delta\pa_kZ_k[J,\Phi^*]=\Big(\frac{i}{\hbar}\Big)^2
\pa_kS_k[-i\hbar\overrightarrow{\pa}_{\!\!J}]
\delta\Psi_{,A}[-i\hbar\overrightarrow{\pa}_{\!\!J}]
R^A(-i\hbar\overrightarrow{\pa}_{\!\!J})Z_k[J,\Phi^*].
\eeq
In terms of the functional $W_k[J,\Phi^*]$ we have
\beq
\label{e20}
\delta\pa_kW_k[J,\Phi^*]=
\pa_kS_k[\overrightarrow{\pa}_{\!\!J}W_k-i\hbar\overrightarrow{\pa}_{\!\!J}]
\delta\Psi_{,A}[\overrightarrow{\pa}_{\!\!J}W_k-i\hbar\overrightarrow{\pa}_{\!\!J}]
R^A(\overrightarrow{\pa}_{\!\!J}W_k-i\hbar\overrightarrow{\pa}_{\!\!J})\cdot 1
\eeq
Finally, the gauge dependence of the flow equation is described by the equation
\beq
\label{e21}
\delta\pa_k\Gamma_k[\Phi,\Phi^*]=
\pa_kS_k[{\hat \Phi}]
\delta\Psi_{,A}[{\hat \Phi}]
R^A({\hat \Phi})\cdot 1 .
\eeq
Therefore, at any finite value of $k$ the  effective average action
depends on gauges.
But what is about the case when $k\rightarrow 0$?
One can think that due to the property
\beq
\label{e22} \lim_{k \rightarrow 0} \Gamma_k=\Gamma,
\eeq where
$\Gamma$
is the standard effective action constructed by the
Faddeev-Popov rules, the gauge dependence of  effective average
action disappears at the fixed points (see, for example, \cite{deAlwis}).
It is not true
because by itself the effective action $\Gamma$ depends on gauges.
Moreover there exists an additional reason to doubt the gauge
independence of effective average action at the fixed points. Indeed,
in the FRG the effective average action $\Gamma_k$ should be found
as a solution to the flow equation (\ref{e17}) which includes the
differential operation with respect to the IR parameter $k$. Let us
present the  effective average action in the form
\beq
\label{e23}
\Gamma_k=\Gamma+k H_k,
\eeq
where functional $H_k$ obeys the
property
\beq
\label{e24} \lim_{k \rightarrow 0}H_k=H_0\neq 0.
\eeq
Then we have the relations
\beq
\label{e25} \pa_k\lim_{k \rightarrow
0} \Gamma_k=0,\quad \lim_{k \rightarrow 0} \pa_k\Gamma_k=H_0.
\eeq
These two operation do not commute and the gauge
independence at the fixed points requires some additional study.
Taking into account the commutativity of gauge variation and of the limit $k\rightarrow 0$
from Eqs. (\ref{e21}) and (\ref{e25}) it follows
\beq
\quad \lim_{k \rightarrow 0} \delta\pa_k\Gamma_k=\delta H_0.
\eeq
If $H_0$ depends on gauges then $\delta H_0\neq 0$ and one meets the gauge
dependence problem at the fixed points.
Existence of this problem we are going to support by explicit calculations of effective
average action for  a toy gauge model  based on  electromagnetic field
in the flat space-time.

The classical action of the model is
 \beq
 \label{e26}
 S_0(A)=-\frac{1}{4} \int d^4 x\, F_{\mu\nu} F^{\mu\nu}\,, \qquad F_{\mu\nu}
 = \pa_\mu
 A_\nu - \pa_\nu A_\mu\,.
 \eeq
We choose the gauge fixing function in the form corresponding to
non-singular gauges
 \beq
 \label{e27}
 \chi(A,B) = \frac{1}{\sqrt{1+\xi}}\, \pa^\alpha A_\alpha + B\,.
 \eeq
where $B$ is an auxiliary field introducing the gauge and $\xi$ is a gauge parameter.
Integrating over field $B$ in the functional integral yields
the gauge fixing action
 \beq
 \label{e28}
 S_{gf}(A) = - \frac{1}{2(1+\xi)} \int d^4 x\, (\pa^\alpha A_\alpha)^2\,.
 \eeq
The action for ghosts reads
 \beq
 \label{e29}
 S_{gh}(\bar C, C) = \frac{1}{\sqrt{1+\xi}} \int d^4 x\, \bar C (\pa^\alpha \pa_\alpha) C\,.
 \eeq

Calculation of the   effective average action of the model within the standard
FRG method gives
 \beq
 \label{e30}
 \Gamma_k(\Phi) = S_0(A)+S_{gf}(A)+S_{gh}(\bar C, C)+ S_k(\Phi)
 + i\hbar\,\Gamma^{(1)}_k(\xi),
\eeq
where the regulator action, $S_k(A,\bar C, C)$, is
\beq
\label{e31}
S_k(\Phi)= \frac{1}{2} \int d^4x\, A^\alpha (R_{k,\,A})_{\alpha\beta}A^\beta
+ \int d^4 x\, \bar C R_{k,\,gh} C,
 \eeq
and  $\Gamma^{(1)}_k(\xi)$ reads
 \beq
 \label{e32}
\Gamma^{(1)}_k(\xi) =
 \frac{1}{2}\Tr \ln\Big(\square \delta^\alpha_\beta
 -\frac{\xi}{1+\xi}\pa^\alpha\pa_\beta + (R_{k,\,A})^\alpha_\beta \bigg)
- \Tr\ln\Big(\frac{1}{\sqrt{1+\xi}}\,\square +  R_{k,\,gh}\bigg)
 \,.
 \eeq
It is important to note that the action (\ref{e30})  is the exact
solution to  the flow equation without using any truncation schemes.

From (\ref{e30})-(\ref{e32}) it follows
\beq
\nonumber
&&\lim_{k\rightarrow 0}\Gamma_k(\Phi) =S_0(A)+S_{gf}(A)+S_{gh}(\bar C, C)+\\
\label{e33}
&&\qquad\qquad\quad+
 i\hbar\,\frac{1}{2}\Tr \ln\Big(\square \delta^\alpha_\beta
 -\frac{\xi}{1+\xi}\pa^\alpha\pa_\beta \Big)
- i\hbar\, \Tr\ln\Big(\frac{1}{\sqrt{1+\xi}}\,\square \Big), \,
\eeq
and
\beq
\nonumber
\pa_k\Gamma_k(\Phi) &=&\pa_kS_k(\Phi)+ i\hbar\,\frac{1}{2}
\Tr \big[G^{\alpha}_{\beta}(\xi)\pa_k(R_{k,\,A})^\beta_\gamma\big]+\\
\label{e34}
&&+ i\hbar\,\Tr\Big[\Big(\frac{1}{\sqrt{1+\xi}}\,\square +  R_{k,\,gh}\bigg)^{-1}
\pa_kR_{k,\,gh}
\Big],
\eeq
where $G^{\alpha}_{\beta}(\xi)$  is an operator inverse  to
\beq
\label{e35}
M^{\alpha}_{\beta}(\xi)=\square \delta^\alpha_\beta
 -\frac{\xi}{1+\xi}\pa^\alpha\pa_\beta + (R_{k,\,A})^\alpha_\beta ,\quad
M^{\alpha}_{\beta}(\xi)G^{\beta}_{\gamma}(\xi)=\delta^{\alpha}_{\gamma}.
\eeq
Therefore the relations (\ref{e33}), (\ref{e34}) confirm main statements about
gauge dependence in the FRG: the effective average action depends on gauges
in the limit $k\rightarrow 0$ and the flow equation depends on gauges at any value of
IR parameter $k$. Moreover, if the partial derivatives of regulator functions
with respect to parameter $k$ do not disappear in the limit  $k \rightarrow 0$,
\beq
\lim_{k\rightarrow 0}\pa_kR_k\neq 0
\eeq
then in this case the second limit in Eq. (\ref{e25}) depends on gauges explicitly.
Let us emphasize again that the toy model is useful in studying basic properties
of effective average action in the FRG due to  the its explicit form of this action.
It allows to analyze the gauge dependence not only the effective average action but the flow
equation at any value of ir parameter.
In particular, this study indicates on existence of a real problem
with  gauge dependence even at the fixed points.

Quite recently by explicit calculations in the FRG approach
the gauge dependence of some mass parameters in gravity theories
at the fixed points has been found \cite{oy}. It means that all general conclusions
made in this subsection about gauge dependence in the FRG are true.

\section{Background field method}
\noindent
The background field method (BFM) \cite{DeW, AFS,Abbott} presents a
reformulation of quantization procedure for Yang-Mills theories
allowing to work with  the effective action invariant under the
gauge transformations of background fields and to reproduce all
usual physical results by choosing a special background field
condition \cite{Weinberg,Abbott}. Application of the BFM
 simplifies essentially calculations of Feynman
diagrams  in gauge theories \cite{K-SZ,GvanNW,CMacL,IO,Gr}
(among recent applications of this
approach see, for example, \cite{Barv,BLT-YM,FrenT, BLT-YM2,BBLT-BV}).
The gauge dependence
problem in this method remains very important matter although it
does not discuss  because standard considerations are restricted by
the background field gauge condition only.

We study the gauge dependence of generating
functionals of Green's functions in the BFM for
Yang-Mills theories in class of gauges depending on gauge and
background vector fields. The background field gauge condition
belongs them as a special choice. We prove that the gauge invariance
can be achieved if the gauge-fixing functions satisfy a tensor
transformation law. We consider the gauge dependence and gauge
invariance problems within the background field formalism as two
independent ones. To support this point of view we analyze the
FRG approach \cite{Wet1,Wet2} in
the BFM. We find restrictions on tensor
structure of the regulator functions which allow to construct a
gauge invariant average effective action. Nevertheless, being gauge
invariant this action remains a gauge-dependent quantity on-shell
making impossible a physical interpretation of results obtained for
gauge theories.

\subsection{BFM in FP-method}

\noindent
We consider any Yang-Mills type of  gauge theory of fields $A^i$, with
Grassmann parity $\vp_i = \vp(A^i)$.  Application of the BFM requires
specifying gauge fields of initial action $S_0[A]$ being invariant under gauge
transformations, $\delta_\xi A^i = R^i_\alpha(A) \xi^\alpha$,
$\vp(R^i_\al) = \vp_i + \vp_\al,\;\vp(\xi^\alpha)=\vp_\alpha.$
A complete set of
fields $\,A^i=(A^{\alpha k},A^m)\,$ includes fields $A^{\al k}$
of the gauge sector and also fields $A^m$ of the matter sector
of a given theory. We do not assume linearity in the fields of
the gauge generators $R^i_\al(A)$ because quite recently generalization
of the BFM for nonlinear gauge-fixing conditions and nonlinear
realizations of the gauge generators has been found \cite{GLSh}.

The BFM story begins with splitting the original
fields $A^i$ into two types of fields, through the substitution
$A^i \longmapsto A^i + \mathcal{B}^i$ in the initial action
$S_0[A]$. It is assumed that the fields $\mathcal{B}^i$ are not equal
to zero {\it only} in the gauge sector. These fields form a classical
background, while $A^i$ are quantum fields, which means
being subject of quantization; \textit{e.g.},
these fields are integration variables
in functional integrals. It  is clear that the total action satisfies
\beq
\label{VarYM-back}
\de_\om S_0 [A+\mathcal{B}] = 0
\eeq
under the transformation
$\,A^i \longmapsto A^{\prime i}
= A^i + R^i_\al (A+\mathcal{B}) \om^\al$.
On the other hand, the new field $\mathcal{B}^i$ introduces extra
new degrees of freedom and, thence, there is an ambiguity in the
transformation rule for each of the fields $A^i$ and $\mathcal{B}^i$.
This ambiguity can be fixed in different ways, and in the BFM, it is
done by choosing the transformation
laws
\beq
\label{BGTrans}
\de_\om^{(q)} A^i
= \left[ R^i_\al (A+\mathcal{B})
- R^i_\al (\mathcal{B}) \right] \om^\al ,
\qquad
\de_\om^{(c)} \mathcal{B}^i = R^i_\al (\mathcal{B}) \om^\al,
\eeq
defining the \textit{background field transformations} for the fields
$A^i$ and $\mathcal{B}^i$, respectively. In linear realization
of gauge generators, the transformations (\ref{BGTrans})
in the sector of fields $A^i$ are just in the form $\de_\om^{(q)} A^i
=R^i_\al (A)\om^\al$.
The superscript ($q$)
indicates the transformation of the quantum fields, while that of the
classical fields is labeled by ($c$). Thus, in
Eq.~\eqref{VarYM-back} one has
$\de_\om = \de^{(q)}_\om + \de^{(c)}_\om$. Indeed, the background
field transformation rule for the field $A^i$ was chosen so that
\beq
\de_\om^{(c)} \mathcal{B}^i + \de_\om^{(q)} A^i
\,=\, R^i_\al (A+\mathcal{B}) \om^\al .
\label{backtrans}
\eeq

Quantization of gauge theory with action $S_0 [A+\mathcal{B}]$
and gauge generators $ R^i_\al (A+\mathcal{B})$ is performed
in the FP method \cite{FP}.
It means that one has to introduce a gauge-fixing condition
for the quantum fields $A^i$ and the set of all quantum fields
$\phi = \left\lbrace \phi^A \right\rbrace $ as described in Sec. 3.
The corresponding Faddeev-Popov action
in the BFM reads
\beq
\label{S-FP_BFM}
S_{\text{FP}} [\phi,\mathcal{B}]
\,=\, S_{0} [A+\mathcal{B}] +
\Psi[\phi,\mathcal{B}]{\hat R}({\phi,\mathcal{B}}),
\eeq
where the notations
\beq
\label{G1}
&&\qquad{\hat R}({\phi,\mathcal{B}})=
\overleftarrow{\pa}_{\!\!\phi^A}R^A(\phi,\mathcal{B}),\quad
\Psi[\phi,\mathcal{B}]=
\bar{C}^\al \chi_\al(A,B,\mathcal{B}),\\
&&
\label{G2}
R^A(\phi,\mathcal{B})=
\big( R^{i}_\al (A+\mathcal{B}) C^\al  , \,\, 0,
\,\,  -(1/2) F^{\al}_{\be\ga} C^\ga C^\be (-1)^{\vp_\be },
\,\,  (-1)^{\vp_\al} B^\al  \big)
\eeq
are used. In (\ref{G1}), $\chi_\al(A,B,\mathcal{B})$ are gauge-fixing functions
which may depend on fields $B^{\alpha}$ allowing us
to introduce nonsingular gauges,
\beq
\label{Gauges-BFM}
\quad
\chi_\al(A,B,\mathcal{B})=\chi_\al(A,\mathcal{B})+
(\xi/2)\; g_{\al\be}B^{\alpha}.
\eeq
In this expression, $\xi$  is a gauge parameter that has to
be introduced in the case of a nonsingular gauge condition, and
$g_{\al\be}$ is an arbitrary invertible constant matrix such that
$g_{\be\al}=g_{\al\be}(-1)^{\vp_\al \vp_\be}$.
 The standard choice of
$\,\chi_\al(A,\mathcal{B})\,$ in the BFM is of the type
$\chi_\al(A,\mathcal{B}) \,=\,F_{\al i}(\mathcal{B})A^i,$
which is a gauge-fixing condition linear in the quantum fields $A^i$.
In what follows, consequent results do not require any kind of \textit{a priori}
specific dependence of the gauge-fixing functions
$\chi_\al(A,B,\mathcal{B})$ on
$A^i, B^\al$ and $\mathcal{B}^i$.

The action (\ref{S-FP_BFM}) is invariant under the BRST transformations
\beq
\label{Btr}
\de_{B}\,\phi^A=R^A(\phi,\mathcal{B}) \mu\;\!,\quad
S_{\text{FP}} [\phi,\mathcal{B}]{\hat R}({\phi,\mathcal{B}})=0,
\eeq
which do not depend on choice of the gauge-fixing condition. In (\ref{Btr}),
 $\mu$ is a constant anticommuting parameter.
The BRST transformations are applied only on quantum fields;
thus, $\de_{B} \, \mathcal{B}^i = 0$. Notice that the BRST operator
is nilpotent,
\beq
\hat{R}^2(\phi,\mathcal{B}) = 0.
\eeq

Apart from the global supersymmetry (BRST symmetry),
a consistent formulation of the
BFM requires that the Faddeev-Popov action be invariant under
background field transformations. The former symmetry is ensured
in the representation~(\ref{S-FP_BFM}) of the Faddeev-Popov action,
for any choice of gauge-fixing functional $\Psi$. Therefore, it is
possible to extend considerations to a more general case in which
$\Psi(\phi,\mathcal{B}) = \bar{C}^\al \chi_\al(\phi,\mathcal{B})$,
where the gauge-fixing functions $\chi_\al(\phi,\mathcal{B})$ depend
on \textit{all} the fields under consideration and satisfy the condition
$\vp(\chi_\al)=\vp_{\al}$. On the other
hand, the presence of the background field symmetry is not immediate
--- especially in the case of nonlinear gauges --- as the gauge-fixing
functionals depend on the background fields. Below, we
derive necessary conditions that the fermion gauge-fixing functional
should satisfy to achieve the consistent application of the BFM.

Let us extend the transformation rule~\eqref{BGTrans} to the whole
set of quantum fields, as
\beq
\label{BGtrans_other}
\quad \de_\om^{(q)} B^\al
= -F^{\al}_{\ga\be}  B^\be \om^\ga ,
\quad
\de_\om^{(q)} C^\al = -F^{\al}_{\ga\be}  C^\be \om^\ga
 (-1)^{\vp_{\ga}} ,
 \quad \de_\om^{(q)} \bar{C}^\al
= -F^{\al}_{\ga\be}  \bar{C}^\be \om^\ga  (-1)^{\vp_{\ga}}.
\mbox{\,\,}
\eeq
Following the procedure used for the BRST symmetry, one can define the
operator of background field transformations,
\beq
\label{R_omega}
\hat{{R}}_\om (\phi,\mathcal{B})=
\overleftarrow{\pa}_{\!\!\mathcal{B}^i}\;\de^{(c)}_\om \mathcal{B}^i
\,+\, \overleftarrow{\pa}_{\!\!\phi^A} \,
\de^{(q)}_\om \phi^A , \qquad \vp(\hat{{R}}_\om) \,=\, 0 .
\eeq
The gauge invariance of the initial classical action implies that
$S_{0} (A+\mathcal{B})\hat{{R}}_\om (\phi,\mathcal{B})  = 0$.
Furthermore, it is not difficult to verify that the background gauge
operator, $\hat{{R}}_\om =\hat{{R}}_\om (\phi,\mathcal{B})$,
commutes with the generator of BRST transformations,
$\hat{R}=\hat{R}(\phi,\mathcal{B})$,
\textit{i.e.},
\beq
[\hat{R}, \hat{{R}}_\om ] = 0.
\eeq
Combining this result with the representation~(\ref{S-FP_BFM}) of the
Faddeev-Popov action, we get
\beq
\label{Condition}
\de_\om S_{\text{FP}}(\phi,\mathcal{B})
=S_{\text{FP}}(\phi,\mathcal{B})\hat{{R}}_\om (\phi,\mathcal{B}) = 0
\quad
\Longleftrightarrow \quad
\, \Psi(\phi,\mathcal{B})\hat{{R}}_\om (\phi,\mathcal{B}) = 0 .
\eeq
In other words, the Faddeev-Popov action is invariant under background
field transformations if and only if the fermion gauge-fixing functional is
a scalar with respect to this transformation.
The condition~\eqref{Condition} constrains the possible forms of
the (extended) gauge-fixing function $\chi_\al(\phi,\mathcal{B})$,
as the relation
\beq
\Psi(\phi,\mathcal{B}) \hat{{R}}_\om (\phi,\mathcal{B})=
\bar{C}^\al \de_\om \chi_\al(\phi,\mathcal{B}) -
F^{\al}_{\ga\be} \bar{C}^\be \om^\ga  (-1)^{\vp_{\ga}}
\chi_\al(\phi,\mathcal{B}) \,=\, 0
\eeq
fixes the transformation law for $\chi_\al(\phi,\mathcal{B})$,
\beq
\label{conditionChi}
\de_\om \chi_\al(\phi,\mathcal{B})
\,=\,  - \chi_\be (\phi,\mathcal{B})  F^{\be}_{\al\ga} \om^\ga.
\eeq
Therefore, to have the invariance of the Faddeev-Popov action
under  background field transformations, it is necessary that the
gauge function $\chi_\al$ transforms as a tensor with respect to the
gauge group.
This requirement can be fulfilled, provided that
$\chi_\al(\phi,\mathcal{B})$ is constructed only by using tensor
quantities.
Thus, Eq.~\eqref{conditionChi} may impose a restriction on the form
of gauge-fixing functions which are nonlinear on the fields $A^i$.
In particular, if the gauge-fixing function $\chi_\al(\phi,\mathcal{B})$
is chosen in a form leading to invariance of the gauge-fixing action under
the background gauge transformations, then the ghost action by itself will be
invariant under these transformations as well.

At this point, we can conclude that (\ref{Btr}) and (\ref{Condition})
represent necessary conditions for the consistent application of the
BFM. The first relation is associated to
the gauge independence of the vacuum functional, which
is needed for the gauge-independent $S$ matrix and hence is a very
important element for the consistent quantum formulation of a gauge
theory \cite{BV1,KT},
while the second relation is called to provide the invariance of the
effective action in the BFM with respect to deformed (in the general
case) background field transformations. In what follows, we shall
consider these statements explicitly. To this end, it is convenient
to introduce the extended action
\beq
S_\text{ext} [\phi,\mathcal{B},\Phi^\ast]
\,=\,
S_\text{FP} [\phi,\mathcal{B}]
+ \Phi^\ast_A \, R^A(\phi,\mathcal{B}) ,
\eeq
where $\Phi^\ast = \lbrace \Phi^{\ast}_{ A}\rbrace$ denote as usual the set
of sources (antifields) to the BRST transformations, with the parities
$\vp(\Phi^{\ast}_{ A})=\vp_A + 1$. The corresponding (extended)
generating functional of Green's functions reads
\beq
\nonumber
Z[J,\mathcal{B},\Phi^\ast]
&= &\int \mathcal{D}\phi \exp\left\lbrace \frac{i}{\hbar}
\big( S_\text{FP} [\phi,\mathcal{B}]  + J_A \phi^A
+ \Phi^\ast_A \, R^A(\phi,\mathcal{B}) \big)  \right\rbrace =\\
\label{Zext}
&=&\exp\left\lbrace \frac{i}{\hbar} W[J,\mathcal{B},\Phi^\ast] \right\rbrace,
\eeq
where
$J_A
= \big( J_i  , \,\, J^{(B)}_{\al}, \,\, \bar{J}_\al , \,\, J_\al \big)$
[with the parities $\vp(J_A)=\vp_A$]
are the external sources for the fields $\phi^A$. The BRST symmetry,
together with the requirement that the generators $R^i_\al$ of gauge
transformation satisfy
\beq
\label{Jaco}
(-1)^{\vp_i} \overrightarrow{\pa}_{\!\!A^i} R^i_\al ( A+\mathcal{B})
+ (-1)^{\vp_\be + 1} F^\be_{\be\al}  = 0\; \Leftrightarrow\;
R^A_{\;,A}(\phi, \mathcal{B})=0,
\eeq
implies in the ST identity
\beq
\label{Slavnov-Taylor}
J_A \overrightarrow{\pa}_{\!\!\Phi^{\ast}_{ A}}
Z[J,\mathcal{B},\Phi^\ast]   \,=\,0.
\eeq
The relation~\eqref{Jaco} plays an important  role in the derivation
of the Ward identity insomuch as it ensures the triviality of the
Berezenian related to the change of integration variables in the form
of BRST transformations.

In terms of the generating functional $W[J,\mathcal{B},\Phi^\ast]$
of connected Green's functions, the ST identity reads
\beq \label{Slavnov-TaylorW}
J_A \,\overrightarrow{\pa}_{\!\!\Phi^{\ast}_{ A}}
 W[J,\mathcal{B},\Phi^\ast) = 0.
\eeq
The (extended) effective action is defined as
\beq
\Ga \,=\,  \Gamma[\Phi,\mathcal{B},\Phi^\ast] =
W[J,\phi^\ast,\mathcal{B}] - J_A \Phi^A,\quad
\Phi^A=
\overrightarrow{\pa}_{\!\!J_A} W[J,\mathcal{B},\Phi^\ast]
\eeq
and it satisfies the ST  identity
\beq
\Gamma\;\overleftarrow{\pa}_{\!\!\Phi^A}\overrightarrow{\pa}_{\!\!\Phi^*_A}\;
\Gamma  = 0 ,
\eeq
written in the form of Zinn-Justin equation \cite{Z-J}.

Let $Z_{\Psi}[{\cal B}]=Z[0,{\cal B},0]$ be the vacuum functional
 which corresponds
to the choice of gauge-fixing functional $\Psi[\phi, {\cal B}]$ in the presence
of external fields ${\cal B}$,
\beq
\label{a17} Z_{\Psi}[{\cal
B}]=\int
D\phi\;\exp\Big\{\frac{i}{\hbar}S_{FP}[\phi, {\cal B}]\Big\}.
\eeq
In turn, let $Z_{\Psi+\delta\Psi}$ be the vacuum functional
corresponding to a gauge-fixing functional $\Psi[\phi,{\cal
B}]+\delta\Psi[\phi,{\cal B}]$,
\beq
\label{a18}
Z_{\Psi+\delta\Psi}[{\cal B}]=\int
d\phi\;\exp\Big\{\frac{i}{\hbar}\big(S_{FP}[\phi, {\cal
B}]+\delta\Psi[\phi,{\cal B}]{\hat R}(\phi,{\cal B})\big)\Big\}.
\eeq
Here, $\delta\Psi[\phi,{\cal B}]$ is an arbitrary infinitesimal
odd functional which  may  in general have  a form  differing from
(\ref{G1}). Making use of the change of variables $\phi^i$ in the
form of BRST transformations but with replacement of the
constant parameter $\mu$ by the functional
\beq
\label{a19}
\mu=\mu[\phi,{\cal
B}]=\frac{i}{\hbar}\delta\Psi[\phi,{\cal B}]
\eeq
and taking into
account that the Jacobian of transformations is equal to
\beq
\label{a20}
J=\exp\{-\mu[\phi,{\cal B}]{\hat R}(\phi,{\cal B})\},
\eeq
we find the gauge independence of the vacuum functional
\beq
\label{a21}
Z_{\Psi}[{\cal B}]=Z_{\Psi+\delta\Psi}[{\cal B}].
\eeq
The property (\ref{a21}) was a reason to omit the label $\Psi$ in
the definition of generating functionals (\ref{Zext}), and it
 means that, due to the equivalence
theorem \cite{KT}, the physical $S$ matrix does not depend on the
gauge fixing.

The vacuum functional $Z[{\cal B}]=Z_{\Psi}[{\cal B}]$ obeys the very
important property of gauge invariance with respect to gauge
transformations of external fields,
\beq
\label{a26}
\delta_{\omega}^{(c)}{\cal B}^i=R^i_{\alpha}({\cal B})\omega^{\alpha},
\quad \delta^{(c)}_{\omega}Z[{\cal B}]=0.
\eeq
It means
the gauge invariance of functional $W[{\cal B}]=W[0, {\cal B},0]$,
$\delta^{(c)}_{\omega}W[{\cal B}]=0$, as well.
The proof is based on using
the change of variables $\phi^A\;\rightarrow\;\phi^A+
\delta_{\omega}^{(q)}\phi^A$
in the functional integral (\ref{a17}) where $\delta_{\omega}^{(q)}\phi^A$ are
defined in Eqs. (\ref{BGTrans}), (\ref{BGtrans_other})
and taking into account
that the Jacobian of these transformations  is equal to a unit,
and assuming the transformation law of gauge-fixing functions
$\chi_{\alpha}$ according to $\de_\om \chi_\al(\phi,\mathcal{B})
\,=\,  - \chi_\be (\phi,\mathcal{B})  F^{\be}_{\al\ga} \om^\ga$.
In particular, we can argue the invariance
of $S_{FP}[\phi, {\cal B}]$ under combined gauge transformations of external and
quantum fields
\beq \label{a29}
\delta_{\omega}S_{FP}[\phi, {\cal B}]=0.
\eeq
In its turn from the second in (\ref{a26}) and
the relation $W[\mathcal{B}]=-i\hbar\ln Z[\mathcal{B}]$,
it follows the invariance of functional
$W[\mathcal{B}]$,
\beq
\delta^{(c)}_{\omega}W[{\cal B}]=0
\eeq
under the background gauge transformations.
Finally, the main object of the BFM, namely, the effective action of background fields,
$\Gamma[{\cal B}]$, is invariant,
\beq
\delta^{(c)}_{\omega}\Gamma[{\cal B}]=0,
\eeq
under the background gauge transformations as well.

The relations
between the standard generating functionals and the analogous
quantities in the background field formalism are established with
modification of gauge functions \cite{Abbott}.
Here, for the sake of completeness, we compare the generating
functionals  in the BFM  and in the traditional
one --- and, ultimately, their relations with $\Ga[\mathcal{B}]$.
To do this, we consider the generating functional of Green's functions, which
corresponds to the standard quantum field theory approach, but in a
 very special gauge fixing,
\beq
\label{Z2}
Z_2[J]=  \int D\phi \exp\left\lbrace
\frac{i}{\hbar} \big( S_{0} [A] +
\Psi[\phi-\mathcal{B},\mathcal{B}]\hat{R}(\phi)
+ J_A \phi^A \big)  \right\rbrace ,
\eeq
where $\hat{R}(\phi)$ is the generator of standard BRST transformations
(\ref{b10}).
In the last expression, all the dependence of the quantity
$Z_2[J]$ on the external field is only through the
gauge-fixing functional. Thus, this functional depends the external
field $\mathcal{B}^i$, but since this dependence is not of the
BFM type, $Z_2[J]$ is nothing else but the
conventional generating functional of Green's functions of the theory,
defined by $S_0$ in a specific $\mathcal{B}^i$-dependent gauge.
One of the consequences is that any kind of physical results does
not depend on  $\mathcal{B}^i$.  The arguments of
$\Psi$ are written explicitly, showing that we assume that $A^i$
only occurs in a specific combination with $\mathcal{B}^i$.
We stress that, being formulated in the traditional way (i.e., not
in the BFM), $\,Z_2[J]\,$  does not impose
any constraint on the linearity of the gauge-fixing fermion $\Psi$
with respect to the quantum field $A^i$.

Making some change of variables in the functional integral, it is
easy to verify that there exists the relation
\beq
Z[J,\mathcal{B}]= Z_2[J] \exp \Big\{ -\frac{
i}{\hbar} J_i \mathcal{B}^i \Big\},
\eeq
where $Z[J,\mathcal{B}]$ is
the functional $Z[J,\mathcal{B},\Phi^*]$ (\ref{Zext}) restricted on
hypersurface $\Phi^*_A=0$. Accordingly, for the generating
functional of connected Green's functions, one has
\beq
\label{W}
W[J,\mathcal{B}] = W_2[J] - J_i \mathcal{B}^i,
\eeq
where $W_2[J] =
-i\hbar \ln Z_2[J]$. Recall that
\beq
\label{MeanField}
{\cal A}^i
= \overrightarrow{\pa}_{\!\!J_i}W[J,\mathcal{B}].
\eeq
Similarly,
\beq
{\cal A}_{2}^i =\overrightarrow{\pa}_{\!\!J_i} W_2[J] = {\cal
A}^i + \mathcal{B}^i .
\eeq
Following the same line, let us define the
effective action associated to $Z_2[J]$, as
\beq
\Ga_2[\Phi_2] = W_2[J] -J_A\Phi^A_2 .
\eeq
A moment's
reflection shows that
\beq
\label{55}
\Ga[\Phi,\mathcal{B}] =\Ga_2[\Phi_2].
\eeq
In other words, the effective action
$\Ga[\Phi,\mathcal{B}]$ in the background field formalism is equal
to the initial effective action in a particular gauge with mean
field $\,{\cal A}_2^i = {\cal A}^i + \mathcal{B}^i$ --- or,
switching off the mean fields,
\beq
\label{56}
\Ga[\mathcal{B}] =
\Ga_2[{\cal A}_2]\big|_{{\cal A}_2=\mathcal{B}}.
\eeq
We
point out that the gauge is not
associated to its linearity with respect to the quantum fields but
to its dependence on the background field [see Eq.~\eqref{Z2}].

Quantization of the Yang-Mills type of gauge theories in the BFM within the
FP method provides  with very attractive features, namely, the BRST symmetry of the
FP action, the background gauge invariance of effective action, and
gauge independence of S-matrix elements.

\subsection{BFM in FRG}

Here, we discuss the background gauge invariance  and gauge dependence
of average effective
action as well as violation of the BRST symmetry in the FRG
\cite{Wet1,Wet2} using the BFM.
Of course, as to the background field symmetry,
this issue is not new ,(see, for example,
\cite{RW1,Fre-Lit-Paw,Wet-2018}), but we are going to remind the reader of the  main results
related to specific  features of the FRG approach in the BFM.
We pay  special attention to the problem of gauge dependence
of the flow equation
as a new issue in our studies of the FRG.

Inclusion of the FRG in the BFM may be achieved in two ways
with the help of special
dependence of regulator functions on background fields \cite{RW1,LRNSh}
when the regulator action
$S_k[\phi, {\cal B}]$ depends on background fields ${\cal B}$ or
due to special tensor
structure of regulator functions \cite{L} when the regulator action $S_k[\phi]$
does not depend on ${\cal B}$. In both realizations, the regulator action $S_k$
is invariant under background gauge transformations
$\delta^{(q)}_{\omega}\phi^A=R^A_{\omega}(\phi,{\cal B})$,
$\delta^{(c)}_{\omega}{\cal B}^i=R^i_{\omega}({\cal B})$
[see the relations (\ref{BGTrans}) and (\ref{BGtrans_other})],
\beq
\label{w1}
\delta_{\omega}S_k=0.
\eeq
In what follows, we use the notation $S_k[\phi, {\cal B}]$ for definiteness.
The full action of the FRG approach in the BFM has the form
\beq
\label{w2}
S_{Wk}[\phi, {\cal B}]=S_{FP}[\phi, {\cal B}]+S_k[\phi, {\cal B}],
\eeq
and is invariant under background gauge transformations,
\beq
\label{w3}
\delta_{\omega}S_{Wk}[\phi, {\cal B}]=0.
\eeq
Consider the generating functional of Green's functions,
\beq
\label{w4}
Z_k[J,{\cal B}]=\int D\phi \,\exp \Big\{\frac{i}{\hbar}
\big[S_{Wk}[\phi,{\cal B}]
+J_A\phi^A\big]\Big\}=\exp \Big\{\frac{i}{\hbar}
W_{k}[J,{\cal B}]\Big\},
\eeq
and variation of this functional with respect to
background gauge transformations of external fields ${\cal B}^i$. We have
\beq
\label{w5}
\delta^{(c)}_{\omega}Z_k[J,{\cal B}]=\frac{i}{\hbar}
\int D\phi \,\delta^{(c)}_{\omega}S_{Wk}[\phi,{\cal B}]
\exp \Big\{\frac{i}{\hbar}
\big[S_{Wk}[\phi,{\cal B}]
+J_A\phi^A\big]\Big\}.
\eeq
Making use the change of integration variables $\phi^A$
in the form of background gauge transformation in the functional
integral (\ref{w5}) and taking into account the invariance of
$S_{Wk}[\phi,{\cal B}]$ (\ref{w3}), we obtain
\beq
\label{w6}
\delta^{(c)}_{\omega}Z_k[J,{\cal B}]=\frac{i}{\hbar}
J_AR^A_{\omega}\big(-i\hbar\overrightarrow{\pa}_{\!\!J}, {\cal B}\big)
Z_k[J,{\cal B}].
\eeq
In terms of generating functional of the connected Green's functions
$W_{k}[J,{\cal B}]$, the relation (\ref{w6}) is rewritten as
\beq
\label{w7}
\delta^{(c)}_{\omega}W_k[J,{\cal B}]=
J_AR^A_{\omega}\big(\overrightarrow{\pa}_{\!\!J}W_k-
i\hbar\overrightarrow{\pa}_{\!\!J}, {\cal B}\big)
\cdot 1.
\eeq
Because of  the linearity of generators $R^A_{\omega}(\phi, {\cal B})$
with respect to $\phi$, we have
\beq
\label{w8}
R^A_{\omega}\big(\overrightarrow{\pa}_{\!\!J}W_k-
i\hbar\overrightarrow{\pa}_{\!\!J}, {\cal B}\big)\cdot 1=
R^A_{\omega}\big(\overrightarrow{\pa}_{\!\!J}W_k, {\cal B}\big),
\eeq
and, therefore,
\beq
\label{w9}
\delta^{(c)}_{\omega}W_k[J,{\cal B}]=
J_AR^A_{\omega}\big(\overrightarrow{\pa}_{\!\!J}W_k,{\cal B}\big).
\eeq
Introducing the  effective average action $\Gamma_k[\Phi,{\cal B}]$ through
the Legendre transformation of $W_k[J,{\cal B}]$,
\beq
\label{w10}
\Gamma_k[\Phi,{\cal B}]=W_k[J,{\cal B}]-J_A\Phi^A,\quad
\Phi^A=\overrightarrow{\pa}_{\!\!J_A}W_k[J,{\cal B}],\quad
\Gamma[\Phi,{\cal B}]\overleftarrow{\pa}_{\!\!\Phi^A}=-J_A,
\eeq
from (\ref{w9}), it follows
\beq
\label{w11}
\delta^{(c)}_{\omega}\Gamma_k[\Phi,{\cal B}]=-
\Gamma[\Phi,{\cal B}]\overleftarrow{\pa}_{\!\!\Phi^A}
R^A_{\omega}(\Phi,{\cal B}),
\eeq
or
\beq
\label{w12}
\delta_{\omega}\Gamma_k[\Phi,{\cal B}]=0.
\eeq
The effective average action $\Gamma[\Phi,{\cal B}]$ is gauge invariant under
the background gauge transformations of all fields $\Phi^A$, ${\cal B}^i$.
In particular, the functional $\Gamma_k[{\cal B}]=
\Gamma_k[\Phi,{\cal B}]\big|_{\Phi=0}$,
\beq
\label{w13}
\delta^{(c)}_{\omega}\Gamma_k[{\cal B}]=0,
\eeq
is invariant under the gauge transformations of external fields ${\cal B}^i$.

The BRST symmetry is broken on the level of action
$\delta_B S_{Wk}[\phi, {\cal B}]$ (\ref{w2}),
\beq
\label{w14}
\delta_B S_{Wk}[\phi, {\cal B}]=\delta_BS_{k}[\phi, {\cal B}]\neq 0,
\eeq
On the quantum level, violation of the BRST symmetry leads to gauge dependence
of vacuum functional
\beq
\label{w16}
Z_{k|\Psi}[{\cal B}]=\int D\phi \,\exp \Big\{\frac{i}{\hbar}
S_{Wk}[\phi,{\cal B}]\Big\}.
\eeq
Indeed, consider the vacuum functional corresponding another choice of
gauge-fixing functional, $\Psi[\phi]+\delta\Psi[\phi]$,
\beq
\label{w17}
Z_{k|\Psi+\delta\Psi}[{\cal B}]=
\int D\phi \,\exp \Big\{\frac{i}{\hbar}
\big(S_{Wk}[\phi,{\cal B}]+\delta\Psi_{,A}[\phi,{\cal B}]
R^A(\phi,{\cal B})\big)\Big\}.
\eeq
Making use the change of integration variables $\phi^A$ in the form
of BRST transformations with replacement constant parameter $\mu$ by
functional $\mu[\phi,{\cal B}]$ and choosing this functional in the form
\beq
\label{w18}
\mu[\phi,{\cal B}]=(i/\hbar)\delta\Psi[\phi,{\cal B}],
\eeq
we obtain
\beq
\label{w19}
Z_{k|\Psi+\delta\Psi}[{\cal B}]=
\int D\phi \,\exp \Big\{\frac{i}{\hbar}
\big(S_{Wk}[\phi,{\cal B}]+\delta_BS_{k}[\phi, {\cal B}]\big)\Big\}.
\eeq
We cannot propose any change of variables in the functional integral
(\ref{w19}) to reduce it to $Z_{k|\Psi}[{\cal B}]$. Therefore,
\beq
\label{w20}
Z_{k|\Psi+\delta\Psi}[{\cal B}]\neq Z_{k|\Psi}[{\cal B}],
\eeq
and the vacuum functional of the FRG approach and the S matrix remain
gauge dependent within the
BFM as well.

To discuss the mST identity, it is useful,
as we know from previous investigations,  to introduce
the extended generating functionals of Green's functions
$Z_k[J,{\cal B},\Phi^*]$ and connected Green's functions
$W_{k}[J,{\cal B},\Phi^*]$,
\beq
\nonumber
Z_k[J,{\cal B},\Phi^*]&=&\int D\phi \,\exp \Big\{\frac{i}{\hbar}
\big[S_{Wk}[\phi,{\cal B}]+\Phi^*_AR^A(\phi,{\cal B})
+J_A\phi^A\big]\Big\}=\\
\label{w21}
&=&
\exp \Big\{\frac{i}{\hbar}
W_{k}[J,{\cal B},\Phi^*]\Big\}.
\eeq
Using the change of variables $\phi^A$ in the form of
BRST transformations (\ref{Btr})
and taking into account the BRST invariance of $S_{FP}[\phi, {\cal B}]$,
we obtain
 \beq
 \label{w22}
\big(J_A\overrightarrow{\pa}_{\!\!\Phi^*_A}+
S_{k,A}[-i\hbar\overrightarrow{\pa}_{\!\!J},{\cal B}]
\overrightarrow{\pa}_{\!\!\Phi^*_A}
\big)Z_k[J,{\cal B},\Phi^*]\equiv 0,
 \eeq
which is  the mST identity in the FRG within the BFM
 written for functional $Z_k[J,{\cal B},\Phi^*]$. It is clear that this identity
coincides with the ST identity (\ref{Slavnov-Taylor}) in the limit
$k\rightarrow 0$.
In terms of the extended generating functional of connected Green's functions,
$W_k=W_k[J,{\cal B},\Phi^*]$, the identity
(\ref{w22}) is rewritten as
\beq
 \label{w23}
\big(J_A\overrightarrow{\pa}_{\!\!\Phi^*_A}+
S_{k,A}[(\overrightarrow{\pa}_{\!\!J}W_k)
-i\hbar\overrightarrow{\pa}_{\!\!J},{\cal B}]
\overrightarrow{\pa}_{\!\!\Phi^*_A}
\big)W_k[J,{\cal B},\Phi^*]\equiv 0 .
 \eeq
The extended  effective average action,
$\Gamma_k=\Gamma_k[\Phi,{\cal B},\Phi^*]$,
is defined through the
Legendre transformation of $W_k=W_k[J,{\cal B},\Phi^*]$,
\beq
 \label{w24}
 \Gamma_k[\Phi,{\cal B},\Phi^*]=W_k[J,{\cal B},\Phi^*]-J\Phi,\quad
 \Phi^A=\overrightarrow{\pa}_{\!\!J_A}W_k[J,\Phi^*],\quad
\Gamma_k[\Phi,{\cal B},\Phi^*]\overleftarrow{\pa}_{\!\!\Phi^A}=-J_A .
\eeq
Then, the identity (\ref{w23}) can be presented in terms of $\Gamma_k$ as
\beq
\label{w25}
\Gamma_k\overleftarrow{\pa}_{\!\!\Phi^A}\overrightarrow{\pa}_{\!\!\Phi^*_A}
\Gamma_k-
S_{k,A}[{\hat \Phi},{\cal B}]
\overrightarrow{\pa}_{\!\!\Phi^*_A}\Gamma_k\equiv 0 ,
\eeq
or, using the antibracket,
\beq
\label{w26}
\frac{1}{2}(\Gamma_k,\Gamma_k)-
S_{k,A}[{\hat \Phi},{\cal B}]
\overrightarrow{\pa}_{\!\!\Phi^*_A}\Gamma_k\equiv 0 ,
\eeq
where the notations
\beq
\label{w27}
{\hat \Phi}^A=\Phi^A+ i\hbar(\Gamma^{''-1}_k)^{AB}\,
\overrightarrow{\pa}_{\!\!\Phi^B},\quad
(\Gamma_k^{''})_{AB}=\overrightarrow{\pa}_{\!\!\Phi^A}\Gamma_k
\overleftarrow{\pa}_{\!\!\Phi^B},\quad
\big(\Gamma^{''-1}_k\big)^{AC}\cdot
\big(\Gamma^{''}_k\big)_{CB}\,=\delta^A_{\,B},
\eeq
are used.

The existence of the background mST identity for functional
$\Gamma_k[\Phi,{\cal B},\Phi^*]$ does not lead to a solution of
gauge-dependence problem in the FRG approach at least for any finite value of
ir parameter $k$. The case when $k\rightarrow 0$ requires special studies
of the gauge-dependence problem of the background flow equation. The background
flow equation can be formulated for the extended background
effective average action
$\Gamma_k[\Phi,{\cal B},\Phi^*]$ or for
the background  effective average action $\Gamma_k[\Phi,{\cal B}]$.
In what follows, we
study  the background flow equation  for functional
$\Gamma_k[\Phi,{\cal B}]$
for two reasons. First, this functional is under scrutiny of
the FRG community,
and second,  being invariant under the background gauge transformations
the functional remains  gauge dependent even on shell. In  turn,
it shows once again that gauge-invariance
and gauge-dependence properties in gauge theories should
be considered as independent ones.

The background flow equation for the functional
$Z_k[J,{\cal B}]$,
\beq
\label{w28}
\pa_kZ_k[J,{\cal B}]=\frac{i}{\hbar}
\pa_kS_{k}[-i\hbar\overrightarrow{\pa}_{\!\!J}, {\cal B}]Z_k[J,{\cal B}],
\eeq
and the corresponding equation for the functional $W_k[J,{\cal B}]$,
\beq
\label{w29}
\pa_kW_k[J,{\cal B}]=
\pa_kS_{k}[\overrightarrow{\pa}_{\!\!J}W_k-i\hbar\overrightarrow{\pa}_{\!\!J},
 {\cal B}]\cdot 1
\eeq
follow from (\ref{w4}).
The background  effective average action,
\beq
\label{w30}
\Gamma_k[\Phi,{\cal B}]=W_k[J,{\cal B}]-J_A\Phi^A,\quad
\Phi^A=\overrightarrow{\pa}_{\!\!J_A}W_k[J,{\cal B}],\quad
\Gamma_k[\Phi,{\cal B}]\overleftarrow{\pa}_{\!\!\Phi^A}=-J_A,
\eeq
satisfies the background flow equation
\beq
\label{w31}
\pa_k\Gamma_k[\Phi,{\cal B}]=
\pa_kS_{k}[{\hat \Phi}]\cdot 1,
\eeq
where the functional differential operators ${\hat \Phi}^A$ are defined
in the form of (\ref{d42})
with the functional $\Gamma_k[\Phi,{\cal B}]$.

Derivation of equation describing the gauge dependence of background
flow equations
(\ref{w29}), (\ref{w30}), and(\ref{w31}) is similar to that used
in subsection 5.3.
The results read
\beq
\label{w32}
\delta\pa_kZ_k[J,{\cal B}]=\Big(\frac{i}{\hbar}\Big)^2
\pa_kS_k[-i\hbar\overrightarrow{\pa}_{\!\!J},{\cal B}]
\delta\Psi_{,A}[-i\hbar\overrightarrow{\pa}_{\!\!J},{\cal B}]
R^A(-i\hbar\overrightarrow{\pa}_{\!\!J},{\cal B})Z_k[J,\Phi^*],
\eeq
\beq
\label{w33}
\delta\pa_kW_k[J,{\cal B}]=
\pa_kS_k[\overrightarrow{\pa}_{\!\!J}
W_k-i\hbar\overrightarrow{\pa}_{\!\!J},{\cal B}]
\delta\Psi_{,A}[\overrightarrow{\pa}_{\!\!J}
W_k-i\hbar\overrightarrow{\pa}_{\!\!J},{\cal B}]
R^A(\overrightarrow{\pa}_{\!\!J}
W_k-i\hbar\overrightarrow{\pa}_{\!\!J}, {\cal B})\cdot 1,
\eeq
\beq
\label{w34}
\delta\pa_k\Gamma_k[\Phi,{\cal B}]=
\pa_kS_k[{\hat \Phi},{\cal B}]
\delta\Psi_{,A}[{\hat \Phi},{\cal B}]
R^A({\hat \Phi},{\cal B})\cdot 1 .
\eeq
At any finite value of ir parameter $k$, the background flow equations
(\ref{w29}), (\ref{w30}), and (\ref{w31}) are gauge dependent
(\ref{w32}), (\ref{w33}), and (\ref{w34}).
At the fixed point, the gauge dependence does not disappear
for same reasons which were given
in the end of subsection 5.3.

We see that application of the background field method does not help
to solve the gauge-dependence problem in the FRG because
the BRST symmetry remains broken \cite{L}.

\section{Discussion}

\noindent
In the paper, the basic properties of gauge theories
in the framework of FP method, BV formalism and FRG approach
have been analyzed. It is known that the FP and BV quantizations
are characterized by the BRST symmetry which  governs  gauge independence
of S-matrix elements. In turn, the BRST symmetry is broken in the FRG
approach with
all negative consequences for physical interpretation of results.
One of the goals of this
work was to study
the gauge dependence of the  effective average action as a solution
of the flow equation.
For the first time, the equation describing the gauge dependence of the flow equation
has been explicitly derived.
The gauge dependence of flow equation
at any finite value of the ir parameter $k$ was found.  As for the limit
$k\rightarrow 0$, there
is a strong motivation given in the paper
(see subsection 5.3) about the gauge dependence of  effective average action
at the fixed point. Quite recently, this point of view has been supported
by explicit calculations of some mass parameters in gravity theories
at the fixed points \cite{oy}.

Despite of above  feature, it was shown that the FP method,
the BV formalism, and the FRG approach can be provided
with the ST identity, the Ward identity, and the
mST identity, respectively. It was stressed
that the existence of these identities is a direct consequence of
gauge invariance of the initial classical action of the gauge theory under
consideration. Presentation of these identities is essentially simplified
by using both the extended generating functionals of Green's functions
and the BRST transformations.

It was proven that using
the background field method  the background gauge invariance
of the effective action within the FP and FRG  quantization procedures
can be achieved in nonlinear gauges.
The gauge-dependence problem within the FP and FRG quantizations
in the framework of BFM was studied. Application of the BFM in the case
of the FRG approach did not help in solving the problem of gauge dependence of
S matrix.
Arguments allowing us to state impossibility of gauge independence of
physical results obtained  within the FRG approach were given.

\section*{Acknowledgments}
\noindent
The author thanks I.L. Shapiro and I.V. Tyutin
for useful discussions. My special thanks to J.M. Pawlowski,
the intensive correspondence with whom caused appearance of this paper.
I am grateful to the anonymous referee for the detailed and kind criticism
that contributed to the improvement of the paper.
The work  is supported by Ministry of Education
of the Russian Federation, Project No. FEWF-2020-0003.

\end{document}